\documentclass[12pt,a4paper]{article}

\usepackage{amsmath,amsfonts,latexsym,amssymb}
\usepackage{verbatim,amsthm,curves,graphics}
\usepackage{mathrsfs}

\theoremstyle{definition}
\newtheorem{thm}{Theorem}[section]

\newtheorem{lemma}{Lemma}[section]
\newtheorem{prop}{Proposition}[section]

\newtheorem{defn}{Definition}[section]

\begin{document}

\newcommand{\myt}{F^0}
\newcommand{\mH}{H^0}
\newcommand{\momega}{\omega^0}
\newcommand{\mytau}{\tau^0}
\newcommand{\myteta}{{u}^{0}}
\newcommand{\myT}{T^{0}}
\newcommand{\mrno}{
\stackrel{{\scriptstyle \times}}{{\scriptstyle \times}}}
\newcommand{\mlno}{
\stackrel{{\scriptstyle \times}}{{\scriptstyle \times}}}
\newcommand{\myn}{\stackrel{\circ}{\partial}}
\newcommand{\myI}{\stackrel{\circ}{I}}
\newcommand{\mr}{\mathbb{R}} 
\newcommand{\mz}{\mathbb{Z}} 
\newcommand{\mc}{\mathbb{C}} 
\newcommand{\mh}{\mathbb{H}} 
\newcommand{\ms}{\mathbb{S}} 
\newcommand{\mn}{\mathbb{N}} 
\newcommand{\F}{{\bf F}} 
\newcommand{\G}{{\bf G}} 
\newcommand{\pgator}{/\!\!\! S}
\newcommand{\rP}{\operatorname{P}}
\renewcommand{\i}{{\rm i}}
\newcommand{\car}{\operatorname{CAR}}
\newcommand{\FT}{\operatorname{FT}}  
\newcommand{\rSO}{\operatorname{SO}}      
\newcommand{\rO}{\operatorname{O}}        
\newcommand{\rCliff}{\operatorname{Cliff}}
\newcommand{\WF}{\operatorname{WF}}       
\newcommand{\Pol}{{\rm WF}_{pol}}         
\newcommand{\cO}{\mathcal{O}}             
\newcommand{\cA}{\mathcal{A}}                
\newcommand{\cW}{\mathcal{W}}                
\newcommand{\fF}{\frak{A}}
\newcommand{\mL}{\mathscr{L}}             
\newcommand{\cP}{\mathcal{P}}             
\newcommand{\cC}{\mathcal{C}}             
\newcommand{\cF}{\mathscr{F}}             
\newcommand{\cE}{\mathcal{E}}             
\newcommand{\cH}{\mathscr{H}}
\newcommand{\cK}{\mathcal{K}}
\newcommand{\cM}{\mathscr{M}}
\newcommand{\cT}{\mathscr{T}}
\newcommand{\cN}{\mathscr{N}}
\newcommand{\cD}{\mathcal{D}}             
\newcommand{\cL}{\mathcal{L}}             
\newcommand{\cQ}{\mathscr{Q}}             
\newcommand{\cI}{\mathcal{I}}             
\newcommand{\bS}{\operatorname{S}}
\newcommand{\bU}{\mathcal{U}}
\newcommand{\bL}{\operatorname{OP}}
\newcommand{\glob}{{\textrm{\small global}}}
\newcommand{\loca}{{\textrm{\small local}}}
\newcommand{\singsupp}{\operatorname{singsupp}}
\newcommand{\dom}{\operatorname{dom}}
\newcommand{\clo}{\operatorname{clo}}
\newcommand{\supp}{\operatorname{supp}}
\newcommand{\rd}{{\rm d}}                 
\newcommand{\mslash}{/\!\!\!}             
\newcommand{\slom}{/\!\!\!\omega}         
\newcommand{\dirac}{/\!\!\!\nabla}        
\newcommand{\myid}{\leavevmode\hbox{\rm\small1\kern-3.8pt\normalsize1}}
\newcommand{\esssup}{\operatorname*{ess.sup}}
\newcommand{\ran}{\operatorname{ran}}
\newcommand{\sd}{{\rm sd}}
\newcommand{\reg}{\,{\rm l.n.o.}}
\newcommand{\wx}{{\bf x}}
\newcommand{\wk}{{\bf k}}
\newcommand{\ws}{{\bf s}}
\newcommand{\lno}{:\!}
\newcommand{\rno}{\!:}
\newcommand{\clim}{\operatorname*{coin.lim}}
\newcommand{\mydef}{\stackrel{\textrm{def}}{=}}
\renewcommand{\min}{{\textrm{\small int}\,\mathscr{L}}}
\newcommand{\Exp}{\operatorname{Exp}}
\newcommand{\Ad}{\operatorname{Ad}}
\newcommand{\sa}{\mathfrak{sa} } 
\newcommand{\so}{\mathfrak{so} } 
\renewcommand{\o}{\mathfrak{o} } 
\newcommand{\su}{\mathfrak{su} } 
\newcommand{\sq}{\mathfrak{sq} } 
\renewcommand{\sp}{\mathfrak{sp} } 
\newcommand{\gl}{\mathfrak{gl}} 
\newcommand{\g}{{\bf g} }	
\newcommand{\h}{\mathfrak{h} } 
\newcommand{\f}{\mathfrak{f} } 
\newcommand{\p}{\mathfrak{p} } 
\newcommand{\U}{\operatorname{U}} 
\newcommand{\Z}{\operatorname{Z}} 
\newcommand{\SO}{\operatorname{SO} } 
\newcommand{\SU}{\operatorname{SU} } 
\renewcommand{\O}{\operatorname{O} }
\newcommand{\SP}{\operatorname{Sp} } 
\newcommand{\SL}{\operatorname{SL}} 
\newcommand{\GL}{\operatorname{GL}} 
\newcommand{\Der}{\operatorname{Der}} 
\newcommand{\Str}{\operatorname{Str}} 
\newcommand{\End}{\operatorname{End}} 
\newcommand{\Cl}{\operatorname{Cl}} 
\newcommand{\ds}{\dot{+}}
\renewcommand{\aa}{\alpha}
\renewcommand{\gg}{\gamma}
\renewcommand{\d}{\delta }
\newcommand{\x}{\bar{x} } 
\newcommand{\y}{\bar{y} }
\newcommand{\s}{\bar{s} }
\newcommand{\bv}{\operatorname*{b.v.}}
\newcommand{\ad}{\operatorname{ad}}
\newcommand{\sgn}{\operatorname{sgn}}
\newcommand{\tr}{\operatorname{tr}} 
\newcommand{\per}{\operatorname{per}} 
\newcommand{\Tri}{\operatorname{Tri}}

\author{Stefan Hollands and Robert M. Wald}

\title{Existence of Local Covariant
Time Ordered Products of Quantum Fields in Curved Spacetime}

\maketitle

\begin{abstract}

We establish the existence of local, covariant time ordered products
of local Wick polynomials for a free scalar field in curved
spacetime. Our time ordered products satisfy all of the hypotheses of
our previous uniqueness theorem, so our construction essentially
completes the analysis of the existence, uniqueness, and
renormalizability of the perturbative expansion for nonlinear quantum
field theories in curved spacetime. As a byproduct of our analysis, we
derive a scaling expansion of the time ordered products about the
total diagonal that expresses them as a sum of products of polynomials
in the curvature times Lorentz invariant distributions, plus a
remainder term of arbitrarily low scaling degree.

\end{abstract}

\section{Introduction}

In order to give a perturbative definition of a nonlinear quantum
field theory in a globally hyperbolic, curved spacetime, it is
necessary to define Wick polynomials and their time ordered products
for the corresponding linear (i.e., non-self-interacting) field. In
the case of a scalar field, a construction of these quantities was
given recently by Brunetti, Fredenhagen and K\"ohler \cite{bfk} and by
Brunetti and Fredenhagen \cite{bf}. However, these authors did not
impose a locality or covariance condition on the Wick polynomials or
their time ordered products. In fact, the Wick polynomials were
constructed in \cite{bfk} by means of a normal ordering prescription
with respect to an arbitrarily chosen Hadamard vacuum state. The Wick
polynomials defined in this manner thereby possess an undesireable
nonlocal dependence upon the choice of this vacuum state. Since no locality
or covariance condition was imposed on the construction of
time ordered-products of Wick polynomials in \cite{bf}---and, indeed,
such conditions could not have been imposed since the Wick polynomials
used in \cite{bf} were not local, covariant fields---the
renormalization ambiguities were found to involve coupling {\it
functions} rather than coupling constants.

In a recent paper \cite{hw}, we introduced the notion of a {\it local,
covariant quantum field}\footnote{
A general notion of local, covariant quantum fields has been given 
by~\cite{bfv}.
}, and we then imposed the requirement that the
Wick polynomials and their time ordered products be local, covariant
quantum fields. We also required that these quantities have a suitable
continuous and analytic dependence upon the spacetime metric and have
a suitable scaling behavior under scalings of the metric. (These
latter notions are well defined only for local, covariant fields.) In
addition, we required the Wick polynomials and their time ordered
products to satisfy various additional properties, namely suitable
commutation relations with the free field, a microlocal spectral
condition, and (for the time ordered products) causal factorization
and unitarity conditions. We refer the reader to \cite{hw} for the
precise statements of all of our conditions as well as a complete
explanation of the algebraic framework within which our conditions
were formulated.

In \cite{hw}, uniqueness theorems were proven for both the Wick
polynomials and their time ordered products. For the Wick polynomials,
we showed that any two constructions that satisfy all of the above
conditions can differ at most by a suitable sum of products of curvature
terms of a definite scaling dimension multiplied by lower order Wick
polynomials of the appropriate dimension. In particular, this implies
that the ambiguity in defining Wick polynomials up to a given finite
order is uniquely characterized by only a finite number of parameters.
A similar uniqueness result was obtained for the time ordered
products, thereby establishing that the ambiguity in defining these
quantities up to any given finite order also is characterized by only
a finite number of parameters. We then showed that $\lambda
\varphi^4$-theory in curved spacetime is renormalizable in the sense that
the ambiguities arising in the perturbative definition of this theory
correspond precisely to the (finite number of) parameters appearing in
the classical Lagrangian (provided that the possible curvature
couplings of the appropriate dimension are included in this
Lagrangian). Again, we refer the reader to \cite{hw} for the precise
statements and proofs of these results.

The above uniqueness theorems, of course, do not address the issue of
whether there actually exists a construction of Wick polynomials and
their time ordered products that satisfies all of our requirements. As
already noted above, in \cite{bfk} Wick polynomials were constructed
via a normal ordering prescription, but they fail to satisfy our
requirement of being local and covariant. However, this deficiency can
be repaired in a relatively straightforward manner by replacing the
normal ordering prescription with respect to a (nonlocally defined)
Hadamard vacuum state by a point-splitting prescription based upon a
locally and covariantly defined Hadamard parametrix. It was proven in
\cite{hw} that such a construction does indeed satisfy all of our
requirements, thus establishing the existence of local, covariant Wick
polynomials.

One might hope that the construction of time ordered products given in
\cite{bf} could be similarly modified to yield local, covariant fields
that satisfy all of our requirements. However, it is not at all
obvious how to do this. As in \cite{bf} (and as will be explicitly
seen in subsection 3.1 below), the essential difficulty in defining
time ordered products arises from the extension of certain
multivariable distributions to the total diagonal; it is here that
regularization/renormalization is needed. The usual momentum space
methods of regularization are inapplicable in a curved, Lorentzian
spacetime, but the Epstein-Glaser prescription \cite{eg} is well defined
\cite{bf}. However, this prescription involves the modification of
test functions by the subtraction of their truncated Taylor series
multiplied by a ``cutoff function''. The introduction of such a cutoff
function makes the prescription inherently nonlocal. Consequently, the
time ordered products defined by this prescription will fail to be
local, covariant fields.

A similar difficulty with the Epstein-Glaser prescription occurs in
Minkowski spacetime, where the introduction of the cutoff function
makes the prescription fail to be Lorentz invariant. However, in
Minkowski spacetime, a cohomology argument can then be used to
establish existence of a satisfactory Lorentz invariant prescription \cite{pr}.
We have not been able to generalize this argument to curved
spacetime. For this reason, the issue of existence of local covariant
time ordered products was left open in \cite{hw}.

The main purpose of this paper is to prove the existence of local
covariant time ordered products, thereby essentially
completing\footnote{\label{der} As pointed out in \cite{hw}, when
defining Wick polynomials involving derivatives of the field, it is
natural to require the vanishing of any Wick product that contains a
factor of the wave operator applied to the field. This requirement was
not imposed in \cite{hw}, so the issue of existence of Wick
polynomials that satisfy this additional condition remains open. 
We are currently investigating this issue; see also~\cite{mor2}.} 
the perturbative construction
of nonlinear quantum field theory in curved spacetime. The basic idea
of our construction is as follows. As already indicated above, our
task is to extend certain distributions on $M^{n+1} \setminus \Delta_{n+1}$
to all of $M^{n+1}$ in a local, covariant manner,
where $M$ denotes the spacetime manifold, $M^{n+1} = \times^{n+1} M$,
and $\Delta_{n+1}$ denotes the total diagonal of $M^{n+1}$,
\begin{equation} 
\Delta_{n+1} = \{(x, x, \dots, x) \mid x \in M\}.
\end{equation}
The key idea which enables us to accomplish this is to analyze the
scaling behavior of the unextended distributions near the total
diagonal. To do so, we first introduce $n$ ``relative coordinates''
$y$ and show that we can view each unextended distribution as a
distribution in $y$ for each fixed $x \in M$ (i.e., our distribution
in $(n+1)$ variables can be viewed as a distribution in the $n$
relative coordinates that is parametrized by the point $x$ on the
total diagonal).  We then show that near the total diagonal, each
unextended distribution in question can be written as a finite sum of
terms together with a ``remainder term'' with the following
properties: (i) The terms in the finite sum are products of curvature
terms in $x$ times distributions, $u$, in $y$ that correspond to
Lorentz invariant distributions in Minkowski spacetime\footnote{For
the Feynman propagator and its powers, these terms would correspond to the
momentum space expressions given in \cite{pf}, since each $u$ can be
given a momentum space representation.}. (ii) The remainder term has a
sufficiently low scaling degree under scaling of $y$. The
distributions, $u$, may then be extended to the total diagonal by
Minkowski spacetime methods, whereas the remainder term can be
extended to the total diagonal by continuity. The resulting extended
distributions can then be shown to provide a definition of local,
covariant time ordered products that satisfy all of our requirements.

The paper is organized as follows. In section 2, we review our
requirements on the definition of time ordered products. These
requirements are the ones previously given in \cite{hw} except that we
have replaced the continuity requirement of \cite{hw} under smooth
variations of the metric with a smoothness requirement. Further
discussion of our new smoothness requirement is given in Appendix A.

In section 3, we reduce the problem of constructing time ordered
products to that of extending certain scalar distributions to the
total diagonal.  In subsection 3.1, we proceed inductively in the
number, $n$, of variables, and reduce the problem to the extension of
the time ordered products in $n+1$ variables to the total diagonal. In
section 3.2, we use a local, covariant version of the Wick expansion
to express these time ordered products in $n+1$ variables as sums of
local Wick products times ``c-number'' distributions, $t^0$. In
subsection 3.3, we then translate our requirements on the definition
of time ordered products into requirements on the extensions of the
distributions, $t^0$, to the total diagonal.

Section 4 is devoted to obtaining the desired extension of $t^0$. In
subsection 4.1, we introduce ``relative coordinates'', $y$, and then
derive our scaling expansion of $t^0$ with the properties indicated
above. (Some properties of the distributions occurring in the scaling
expansion are obtained in Appendix B.) The scaling expansion is then
used to extend $t^0$ in subsection 4.2. Finally, in subsection 4.3, we
show that the extended distributions, $t$, satisfy the properties
listed in subsection 3.3, so that they define a notion of time ordered
products satisfying all of the requirements of section 2. Some
concluding remarks are given in section 5.

We will restrict consideration here to the theory of a scalar field
$\varphi$, but our basic methods and results should be applicable to
other fields. As in \cite{hw}, for notational simplicity we restrict
attention to time ordered products of Wick powers that do not contain
derivatives $\varphi$. However, our results should extend
straightforwardly to time ordered products involving derivatives of
the field, subject to the caveat mentioned in footnote \ref{der}
above.  In addition, for notational simplicity in treating the scaling
behaviour, we restrict consideration to the massless case, so that the
free theory contains no dimensional parameters. Again, our results can
be straightforwardly generalized to the case where dimensional
parameters are present.

\paragraph{Notation and Conventions.}
Our notation and conventions are the same as in \cite{hw}.  In
particular, we define the Fourier transform on $\mr^m$ by $\widehat
u(k) = (2\pi)^{-m/2} \int u(x) e^{+\i kx} d^m x$.  Multi-indices are
denoted by $\alpha = (\alpha_1, \dots, \alpha_m) \in \mn^m_0$. If
$\alpha$ is an $m$-dimensional multi-index, then we also use 
standard notations such as $|\alpha| = \sum \alpha_i$, $x^\alpha =
x_1^{\alpha_1} \dots x_m^{\alpha_m}$ and $\partial^\alpha =
\frac{\partial^{|\alpha|}}{\partial x_1^{\alpha_1} \dots \partial
x_m^{\alpha_m}}$. We also use the ``constant convention'', meaning
that we use the same symbol $C$ for possibly different numerical
constants in a chain of inequalities. The space of compactly supported
smooth functions on a space $X$ with values in the complex numbers is
denoted by $\cD(X)$ and the space of smooth functions on $X$ (not necessarily
of compact support) by $\mathcal{E}(X)$. (For the definition of the 
topology on these spaces, see e.g. \cite[Chap. V]{rs}.)
The corresponding topological dual spaces of distributions are denoted by 
$\cD'(X)$ respectively $\mathcal{E}'(X)$. The elements in 
$\mathcal{E}'(X)$ are the distributions of compact support.
The wave front set \cite{h} of a distribution $u$ is denoted
by $\WF(u)$ and its analytic wave front set \cite{h} (see also
Appendix A) is denoted by $\WF_A(u)$.

\section{Required Properties of the Time Ordered Products}

For the theory of a free scalar field, $\varphi(x)$, on an arbitrary
globally hyperbolic spacetime, $(M, \g)$, we previously defined
\cite{hw} an ``extended Wick-polynomial algebra'', $\cW(M, \g)$, 
which generalizes the construction of D\"utsch and Fredenhagen \cite{df}
to curved spacetimes. This
algebra is sufficiently large to contain elements corresponding to all
Wick powers, $\varphi^k(x)$, (as distributions on compactly supported
test functions on $M$) and their time ordered products
\begin{equation}
\label{top}
T = T(\varphi^{k_1} (x_1) \dots \varphi^{k_n} (x_n)),
\end{equation}
(as distributions on compactly supported smooth test function on
$M^n$). In \cite{hw} we imposed a set of requirements on both
$\varphi^k$ and $T$ that uniquely determined these quantities up to
certain well specified renormalization ambiguities. In \cite{hw}, we
also constructed Wick products satsifying all of our conditions, so in
this paper we will view these quantities as known. Our task here is to
construct time ordered products of Wick powers that satisfy the
following list of requirements, which---apart from the smoothness
condition T4 (see remark (1) at the end of this section)---correspond to
the requirements previously given in \cite{hw}:

\paragraph{T1 Locality/Covariance.}
The time ordered products are local, covariant fields, as defined in
\cite{hw}.

\paragraph{T2 Scaling}
The time ordered products scale ``almost homogeneously'' under rescalings
$\g \to \lambda^{-2} \g$ of the spacetime metric in the following
sense. Let $\Phi$ be a local, covariant field in $n$ variables, 
and let ${\mathcal S}_\lambda \Phi$
be the rescaled local, covariant field given by 
${\mathcal S}_\lambda \Phi[\g] \equiv \lambda^{-4n} \sigma_\lambda 
\Phi[\lambda^{-2} \g]$, where $\sigma_\lambda: \mathcal{W}(M, \lambda^{-2}\g)
\to \mathcal{W}(M, \g)$ is the canonical isomorphism defined in 
\cite{hw}. The scaling dimension, $d_\Phi$, of a local covariant field is 
defined as 
\begin{equation}
d_\Phi = \sup \{ \delta \in \mr \mid \lim_{\lambda \to 0+} 
\lambda^{-\delta} {\mathcal S}_\lambda \Phi = 0\}.
\end{equation}
The scaling requirement on the time ordered product is then 
that 
\begin{equation}
\lambda^{-d_T} {\mathcal S}_\lambda T = 
T + \sum_{h=1}^{N} \ln^h  \lambda \, \Psi_h, 
\end{equation}
where $d_T = \sum k_i$, $N$ is some natural number and where $\Psi_h$ are 
local, covariant fields with scaling dimension $d_T$ which have
fewer powers in the free field than $T$.

\paragraph{T3 Microlocal Spectrum condition.}
Let $\omega$ be any continuous state on $\mathcal{W}(M, \g)$,
so that, as shown in \cite{hr}, $\omega$ has 
smooth truncated $n$-point functions 
for $n \neq 2$ and a two-point function $\omega_2(x, y) 
= \omega(\varphi(x) \varphi(y))$ of Hadamard from, i.e., $\WF(\omega_2)
\subset {\mathcal C}_+(M, \g)$, where 
\begin{equation}
\label{hadadef}
{\mathcal C}_+(M, \g) = \{(x_1, k_1; x_2, -k_2) \in T^* M^2 \setminus
\{0\} \mid (x_1, k_1) \sim (x_2, k_2); k_1 \in (V^+)_{x_1}\}. 
\end{equation}
Here the notation $(x_1, k_1) \sim (x_2, k_2)$ means that $x_1$ and
$x_2$ can be joined by a null-geodesic and that $k_1$ and $k_2$ are
cotangent and coparallel to that null-geodesic. 
$(V^+)_x$ is the future lightcone at $x$. Furthermore, let 
\begin{equation}
\label{gamphid}
\omega_T(x_1, \dots, x_n) = \omega(T(\prod_{i=1}^n \varphi^{k_i}(x_i))).
\end{equation}
Then we require that 
\begin{equation}
\WF(\omega_T) \subset \cC_T(M, \g), 
\end{equation}
where the set $\cC_T(M, \g) \subset T^*M^n \setminus \{0\}$ 
is described as follows (we 
use the graphological notation introduced in \cite{bfk,bf}):  
Let $G(p)$ be a ``decorated embedded graph''
in $(M, \g)$. By this we mean an embedded graph $\subset M$ whose 
vertices are points $x_1, \dots, x_n \in M$
and whose edges, $e$, are oriented null-geodesic curves. Each such null 
geodesic is equipped with a coparallel, cotangent covectorfield $p_e$.  
If $e$ is an edge in $G(p)$ connecting the points $x_i$ and $x_j$ 
with $i < j$, then $s(e) = i$ is its source 
and $t(e) = j$ its target. It is required that
$p_e$ is future/past directed if $x_{s(e)} \notin J^\pm(x_{t(e)})$.
With this notation, we define
\begin{eqnarray}
\label{gamtdef}
\cC_T(M, \g) &=& 
\Big\{(x_1, k_1; \dots; x_n, k_n) \in T^*M^n \setminus \{0\} \mid 
\exists \,\, \text{decorated graph $G(p)$ with vertices} \nonumber\\
&& \text{$x_1, \dots, x_n$ such that
$k_i = \sum_{e: s(e) = i} p_e - \sum_{e: t(e) = i} p_e 
\quad \forall i$} \Big\}. 
\end{eqnarray}

\paragraph{T4 Smoothness.}
The functional dependence of the time ordered products
on the spacetime metric, $\g$, is such that 
if the metric is varied smoothly, then the time ordered
products vary smoothly, in the following sense. 
Consider a smooth one parameter family of metrics $\g^{(s)}$, let $T^{(s)}$
be a corresponding family of time ordered products, and let 
$\cC^{(s)}_T$ be given by eq.~\eqref{gamtdef} 
for this family of metrics. Furthermore, let $\omega^{(s)}$ be a
family of Hadamard states with smooth truncated $n$-point 
functions ($n \neq 2$) depending smoothly on $s$ and with two-point
functions $\omega^{(s)}_2$ depending smoothly on $s$ in the sense 
that (see Appendix A)
\begin{equation}
\WF(\omega_2) \subset \Big\{(s, \rho; x_1, k_1; x_2, k_2) 
\in T^*(\mr \times M^2) \setminus \{0\} \,\Big|\, (x_1, k_1; x_2, k_2) 
\in {\mathcal C}_+^{(s)} \Big\}, 
\end{equation}
where the family of cones $\cC^{(s)}_+$ is defined by 
eq.~\eqref{hadadef} in terms of the family $\g^{(s)}$.
Then we require that the family of distributions given by 
\begin{equation}
\label{family}
\omega_T(s, x_1, \dots x_n) = \omega^{(s)}
(T^{(s)}(\prod_{i=1}^n \varphi^{k_i}(x_i))) 
\end{equation}
depends smoothly on $s$ with respect to 
$\cC^{(s)}_T$ in the sense that 
\begin{equation}
\WF(\omega_T) \subset \Big\{(s, \rho; x_1, k_1; \dots; x_n, k_n) 
\in T^*(\mr \times M^n) 
\setminus \{0\} \,\Big| \,(x_1, k_1; \dots; x_n, k_n) \in \cC_T^{(s)} \Big\}.
\end{equation}

\paragraph{T5 Analyticity.}
Similarly, we require that, for an analytic one-parameter family of analytic
metrics, the expectation value of the time ordered products in an
analytic family of states varies analytically in the same sense as in
T4, but with the smooth wave front set replaced by the analytic wave
front set.

\paragraph{T6 Symmetry.}
The time ordered products are symmetric under a permutation of 
the factors. 

\paragraph{T7 Unitarity.} 
We have $T^* = \bar T$, where $\bar T$ is the ``anti-time-ordered''
product, defined as
\begin{equation}
\bar T(\varphi^{k_1}(x_1) \dots \varphi^{k_n}(x_n)) = 
\sum_{I_1 \sqcup \dots \sqcup I_j = \{1, \dots, n\}}
(-1)^{n + j} T( 
\prod_{i \in I_1} \varphi^{k_i}(x_i)) \dots
T(\prod_{i \in I_j} \varphi^{k_i}(x_i)), 
\end{equation}
where the sum runs over all partitions of the set $\{1, \dots, n\}$ into
disjoint subsets $I_1, \dots, I_j$.

\paragraph{T8 Causal Factorization.}
In the case of a single factor, we require that
$T(\varphi^k(x))=\varphi^k(x)$. For more than one factor, we require
the time ordered product to satisfy the following causal factorization
rule, which reflects the time-ordering of the factors. Consider a
set of points $(x_1, \dots, x_n) \in M^n$ and a partition of $\{1,
\dots, n\}$ into two non-empty disjoint subsets $I$ and $I^c$, with
the property that no point $x_i$ with $i \in I$ is in the past of any
of the points $x_j$ with $j \in I^c$, that is, $x_i \notin J^-(x_j)$
for all $i \in I$ and $j \in I^c$. Then the time ordered products
factorize in the following way:
\begin{eqnarray}
\label{cf}
T = T(\prod_{i\in I}  
\varphi^{k_i}(x_i) )\,T(\prod_{j \in I^c} 
\varphi^{k_j}(x_j)).
\end{eqnarray}

\paragraph{T9 Commutator.}
The commutator of a time ordered product with a free field is given by
lower order time ordered products times suitable commutator functions,
namely 
\begin{equation}
[T(\varphi^{k_1}(x_1) \dots \varphi^{k_n}(x_n)), \varphi(y)] 
=
\i\sum_{i=1}^n k_i \Delta(x_i, y)
T( \varphi^{k_1}(x_1)\dots 
\varphi^{k_i-1}(x_i)  \dots \varphi^{k_n}(x_n)),
\end{equation}
where $\Delta$ is the 
causal propagator (commutator function), defined 
as the difference between the advanced and retarded
fundamental solutions of the Klein-Gordon equation. 

\medskip 

\paragraph{Remarks.}

\noindent
(1) In our paper \cite{hw}, we defined a notion of the continuous
variation of a local covariant field under smooth variations of the
metric, and we imposed this as a requirement on Wick powers and their
time ordered products. We have replaced this requirement here with
condition T4, which requires smooth (rather than continuous)
dependence of the fields. It is easy to verify the the uniqueness
results of \cite{hw} as well as the existence result of \cite{hw} for
Wick powers go through without any essential change if the continuity
requirement imposed there is replaced by condition T4. We prefer to
work with condition T4 here because it is a much simpler condition to
state, it is more general, and it closely parallels the analyticity
requirement T5 that was previously imposed in \cite{hw}. Further
discussion and explanation of conditions T4 and T5 is given in
Appendix A.

\smallskip

\noindent
(2) The microlocal spectrum condition is the same condition as
formulated in \cite{bf}. It may be motivated by the fact that for
noncoinciding points, $\omega(T(\prod \varphi^{k_i}(x_i)))$ can be
expressed in terms of Feynman graphs. A line in such a graph
represents a Feynman propagator, $\omega_F(x, y) \mydef \omega(
T(\varphi(x) \varphi(y))) = 
\omega_2(x,y) - \i\Delta^{\rm adv}(x, y)$, 
whose wave front set off the diagonal is given by \cite{rad}
\begin{equation}
\label{feyn}
\WF(\omega_F) = \{(x_1, k_1; x_2, -k_2) \mid
(x_1, k_1) \sim (x_2, k_2); k_1 \in (V^\pm)_{x_1}  
\Leftrightarrow x_2 \in J^\pm(x_1) \}.   
\end{equation}
For non-coinciding points, the form of $\WF(\omega_T)$ follows from 
\eqref{feyn} and the rules for calculating the wave front set of a product 
of several distributions, see e.g. \cite[Thm. 8.2.10]{h}. 
For coinciding points, the form of $\WF(\omega_T)$
reflects the usual energy momentum conservation rules. On the total diagonal, 
$\Delta_{n}$, the microlocal spectral condition reduces to
\begin{equation}
\WF(\omega_T) \restriction_{\Delta_n}
\perp T(\Delta_n).
\label{perp}  
\end{equation}
where the notation ``$\perp$'' means the following. 
If $F \subset T^*X$ with $X$ a manifold and 
$Y \subset X$ a smooth submanifold, then $F \restriction_Y \perp TY$ 
means that for any $(y, k) \in F \restriction_Y$ and any $(y, v) \in TY$ we 
have that $k_a v^a = 0$.

\smallskip

\noindent
(3) The ``connected time ordered product'', $T^c$, of $n$ 
Wick-monomials is defined in terms of 
the time ordered product by
\begin{eqnarray*}
T^c(\varphi^{k_1}(x_1) \dots \varphi^{k_n}(x_n)) 
&=& \frac{\delta^n}{\i^n \delta f_1(x_1) \dots
\delta f_n(x_n)} \ln S(f) \Big|_{f_1 = \dots = f_n = 0}\\
&=& \sum_{I_1 \sqcup \dots \sqcup I_j = \{1, \dots, n\}}
\frac{(-1)^{j+1}}{j} T( 
\prod_{i \in I_1} \varphi^{k_i}(x_i)) \dots
T(\prod_{i \in I_j} \varphi^{k_i}(x_i)), 
\end{eqnarray*}
where the $f_i$ are test functions of compact support and
$S(f)$ is the formal $S$-matrix for the Lagrangian
$\mL(x) = \sum_i f_i(x) \varphi^{k_i}(x)$,  
\begin{equation}
S(f) = \sum_{n \ge 0} \frac{\i^n}{n!} 
\int_{M^n} T(\mL(x_1) \dots \mL(x_n))
\, \mu_\g(x_1) \dots \mu_\g(x_n).  
\end{equation}
Our unitarity
condition, T9, is equivalent to the condition $T^{c*} = (-1)^{n+1} T^c$ 
on the connected time ordered product. 

\smallskip
\noindent
(4) For Minkowski spacetime, condition T9 was given in \cite{df1,bo}, where
it was shown to be equivalent to the familiar Wick-expansion of the 
time ordered products (see subsection 3.2 below).

\medskip

Our task is to construct time ordered products of Wick powers that
satisfy conditions T1--T9. We shall proceed inductively in the number
of factors, $n$, appearing in the time ordered product (\ref{top}).
By condition T8, for $n=1$ the time ordered products are just the Wick
powers, which were already constructed in \cite{hw}. Therefore, we may
inductively assume that time ordered products with properties T1--T9
have been defined for any number of factors $\le n$. The goal is to
construct from these the time ordered products with $n+1$ factors.  In
the next section, we reduce the problem (in close parallel with the
analysis of \cite{bf}) to that of extending certain multivariable
scalar distributions $t^0$ to the total diagonal.

\section{Reduction to the problem of extending certain scalar 
distributions to the total diagonal} 

\subsection{Construction of time ordered products up to the total diagonal}
 
The key idea of causal perturbation theory is that the time ordered
products with $n+1$ factors are already uniquely determined as
algebra-valued distributions on the manifold $M^{n+1}$ minus its total
diagonal $\Delta_{n+1}$ by the causal factorization requirement T8
(see eq.~\eqref{cf}), once the time ordered products with less than or
equal to $n$ factors are given. Following \cite{bf}, this can be
seen as follows:

Let $I$ be a proper subset of $\{1, 2, \dots, n+1\}$, and let 
$C_I$ be the subset of $M^{n+1}$ defined by 
\begin{equation}\label{ci}
C_I = \{(x_1, x_2, \dots, x_{n+1}) \mid x_i \notin J^+(x_j)
\quad \text{for all $i \in I, j \in I^c$} \}, 
\end{equation}
where $I^c$ is the complement of $I$.
It can be seen that the sets $C_I$ are open and that the collection
$\{C_I\}$ of these sets covers the manifold $M^{n+1} \setminus \Delta_{n+1}$.  
Let $\{f_I\}$ be a
partition of unity subordinate to this covering. On the manifold
$M^{n+1} \setminus \Delta_{n+1}$, we define the algebra-valued
distributions $T^0$ by
\begin{equation}
T^0 = \sum_{I \subsetneq \{1, \dots, n+1\}, I \neq
\emptyset} f_I T_I,  
\label{T0}
\end{equation}
where 
\begin{equation}
T_I = T(\prod_{i \in I} \varphi^{k_i} (x_i)) 
T(\prod_{j \in I^c} \varphi^{k_j} (x_j)).  
\end{equation}
Using the causal factorization property T8 of the time ordered
products with less or equal than $n$ factors, it can be seen that the
definition of $T^0$ does not depend on the choice of the partition
$\{f_I\}$, so $T^0$ is well defined. Property T8 applied to the
time ordered products with $n+1$ factors then requires that the
restriction of $T$ to $M^{n+1} \setminus \Delta_{n+1}$ must agree with
$T^0$. Thus, property T8 alone determines $T$ up to the total diagonal,
as we desired to show.

We now claim that---assuming that time ordered products with less or
equal than $n$ factors have been defined so as to satisfy properties
T1--T9 on $M^{n}$---the fields $T^0$ with $n+1$ factors
automatically satisfy\footnote{Of course, if any $T^0$ failed to
satisfy any of these properties on $M^{n+1} \setminus \Delta_{n+1}$,
we would have a proof that no definition of time ordered products
could exist that satisfies T1--T9.} the restrictions of properties
T1--T9 to $M^{n+1} \setminus \Delta_{n+1}$. Condition T8 can be
immediately seen to hold by virtue of the definition of $T^0$. The
proof that properties T1, T2, T6, T7 and T9 hold is relatively
straightforward. A proof of the microlocal spectral condition, T3, can
be given in exact parallel with reference \cite{bf}. A generalization
of this argument can be used to prove that the smoothness and
analyticity conditions, T4 and T5, also hold.

Our remaining task is to find an extension of each of the
algebra-valued distributions $T^0$ in $n+1$ factors from $M^{n+1}
\setminus \Delta_{n+1}$ to all of $M^{n+1}$ in such a way that properties
T1--T9 continue to hold for the extension. This step, of course,
corresponds to renormalization. Condition T8 does not impose any
additional conditions on the extension, so we need only satisfy
T1--T7 and T9. However, it is not difficult to see that if an
extension $T$ is defined that satisfies T1--T5 and T9, then that
extension can be modified, if necessary, so as to also satisfy the
symmetry and unitarity conditions, T6 and T7. Namely, if the
extension, $T$, of $T^0$ satisfied T1--T5 and T9 but failed to
satisfy the symmetry condition, T6, we could define a new extension
$T'$ by symmetrization, 
\begin{equation}
T' = \frac{1}{(n+1)!}
\sum_{{\rm Perm} \, \pi} T(\varphi^{k_{\pi(1)}}(x_{\pi(1)})
\dots \varphi^{k_{\pi(n+1)}}(x_{\pi(n+1)})).
\end{equation}
The so obtained extension of $T^0$ then satisfies T1--T6 and
T9. Similarly, suppose the extension, satisfied T1--T6 and T9 but
failed to satisfy the unitarity condition, T7, so that the
corresponding connected time ordered product, $T^c$, fails to satisfy
$T^{c*} = (-1)^n T^c$ (see remark (3) of section 2).  Then we define
$T^{c \prime } = \frac{1}{2}( T^c + (-1)^n T^{c*})$ and redefine our
extension by
\begin{equation}
T' = 
T^{c \prime} (\prod_{i=1}^{n+1}\varphi^{k_i}(x_i)) - 
\sum_{
\begin{matrix}
{\scriptstyle
I_1 \sqcup \dots \sqcup I_j = \{1, \dots, n+1\}}\\
{\scriptstyle j \ge 2}
\end{matrix}
}
\frac{(-1)^{j+1}}{j} T( 
\prod_{i \in I_1} \varphi^{k_i}(x_i)) \dots
T(\prod_{i \in I_j} \varphi^{k_i}(x_i)). 
\end{equation}
This provides us with
an extension of $T^0$ that satisfies T1--T7 and T9.

Thus, we have reduced the problem of defining time ordered products to
the problem of extending the distributions $T^0$ defined by
\eqref{T0} from $M^{n+1} \setminus \Delta_{n+1}$ to all of
$M^{n+1}$ so that properties T1--T5 and T9 continue to hold for the
extension. In the next subsection, we will see that property T9 can be
replaced by the requirement of a local Wick expansion for time ordered
products.

\subsection{Reduction to a c-number problem via a local Wick expansion} 

The next key simplification is to reduce the problem of defining the
algebra valued distributions $T$ to the problem of defining certain
``c-number'' distributions $t$. As in \cite{bf}, this is
accomplished by means of a ``Wick expansion''. The usual Wick
expansion in Minkowski spacetime expresses time ordered products as a
sum of normal ordered products with distributional coefficients. In
the generalization to curved spacetime given in \cite{bf}, the
time ordered products are Wick expanded in terms of normal ordered
products defined relative to an arbitrarily chosen quasi-free Hadamard
state. However, such an expansion would not be useful here because the
quasi-free Hadamard state---however it is chosen---has a nonlocal
character. Consequently, the distributional coefficients occurring in
the Wick expansion with respect to normal ordered products will fail
to inherit the locality and covariance properties of the time ordered
products themselves.

For this reason, we will employ here a Wick expansion of the time
ordered procducts with respect to the local, covariant Wick products
$\lno \varphi^{k_1} \dots \varphi^{k_n} \rno_H$ that were previously
defined in \cite{hw} as follows: Let $H(x,y)$ denote the local Hadamard
parametrix
\begin{equation}
\label{hadamp}
H(x, y) = U(x, y) \sigma^{-1} + V(x, y) \ln \sigma, 
\end{equation}
where $U, V$ are certain smooth functions defined in terms of the
metric and the coupling parameters, $\sigma$ is the signed squared
geodesic distance and where the ``$\i 0$'' prescription for the
singular terms is as for the two-point function in Minkowski
space. (The formal power series defining $V$ need not converge in
smooth, non-analytic spacetimes, but a suitably modified convergent
$V$ can be used, as explained in \cite[Sec. 5.2]{hw}.)  Following
\cite[Sec. 5.2]{hw}, we then define in some neighborhood, ${\mathcal
U}_n$, of the total diagonal, $\Delta_n$, the algebra valued
distributions
\begin{eqnarray}
\label{wickexp}
\lno \varphi^{k_1}(x_1)  \dots \varphi^{k_n}(x_n) \rno_H \,\,
= 
\frac{\delta^{|k|}}{\i^{|k|}\delta f(x_1)^{k_1} \dots \delta f(x_{n})^{k_{n}}}
\exp\left(\varphi(f) + \frac{1}{2} H(f,f) \cdot \myid
\right) \Bigg|_{f=0}
\end{eqnarray}
with $|k| = \sum k_i$.
Our Wick-expansion is 
\begin{equation}
\label{wickexp1}
T(\prod_{i=1}^{n} \varphi^{k_i}(x_i) ) = 
\sum_{j_1 \dots j_n} {k \choose j} 
t_{j_1 \dots j_{n}} (x_1, \dots, x_{n})
\lno \varphi^{k_1-j_1}(x_1) \dots  \varphi^{k_n-j_n}(x_n) 
\rno_H,  
\end{equation}
where $t_{j_1 \dots j_{n}}$ are c-number distributions on $\mathcal
U_n$ and where ${k \choose j} = \prod {k_i \choose j_i}$. 
Note that the Wick-expansion formula 
\eqref{wickexp1} is a different identity for different sets of 
exponents $(k_1, \dots, k_n)$, but that the {\it same} coefficients 
$t_{j_1 \dots j_n}$ appear in each identity. 

We emphasize that since our local Wick-products \eqref{wickexp} are
defined only on a sufficiently small neighborhood, $\mathcal U_n$, of
the total diagonal, our Wick-expansion will only make sense in this
neighborhood. (This is in contrast with the Wick-expansion used in
\cite{bf} which is based on a normal ordering prescription for
Wick-products and therefore makes sense everywhere on $M^n$.) This
fact, however, will not cause any complications for our constructions,
since we will need the Wick-expansion only for the purpose of
extending $T^0$ to the total diagonal.

We claim now that any definition of time ordered products that
satisfies requirements T3 and T9 must admit a Wick expansion of the
form \eqref{wickexp1}, with distributional coefficients 
satisfying
\begin{equation}
\label{twf}
\WF(t_{j_1 \dots j_n}) \subset \cC_T(M, \g), 
\end{equation}
where $\cC_T(M, \g)$ is the set specified in \eqref{gamtdef}.  (Note
that \eqref{twf} implies in particular that the products of
distributions implicit in our Wick-expansion formula actually exist
and that the operator given by this formula
defines---after smearing with a smooth test function---an element of
our algebra $\mathcal{W}(M, \g)$.) To prove this claim, we note that
eqs.~\eqref{wickexp1} and~\eqref{twf} 
hold trivially for the time ordered product
$T(\varphi)$ of a single free field. Let us now inductively assume
that eqs.~\eqref{wickexp1} and~\eqref{twf}
have been demonstrated for all time ordered
products of the form $T(\varphi^{k_1} \dots \varphi^{k_n})$, whenever
$|k| = \sum k_i < d$ for some $d \ge 1$. We claim that they also hold for all
multi-orders $(k_1, \dots, k_n)$ with $\sum k_i = d$. To see this, we
consider the difference, 
\begin{multline}
\label{Ddef}
D(x_1, \dots, x_n) = T(\prod_{i=1}^{n} \varphi^{k_i}(x_i) ) -\\ 
\sum_{j_1 \dots j_n, |j| < |k|} {k \choose j} 
t_{j_1 \dots j_{n}} (x_1, \dots, x_{n})
\lno \varphi^{k_1-j_1}(x_1) \dots  \varphi^{k_n-j_n}(x_n) 
\rno_H,  
\end{multline}  
between the left side of eq.~\eqref{wickexp1}
and the expression on the right side of that equation, but with 
the term $t_{k_1 \dots k_n} \, \myid$ omitted in the sum. (Note that this 
is precisely the term in eq.~\eqref{wickexp1} which is not already 
known by the induction hypothesis.)
We now commute $D$ with a free field $\varphi$. We use T9 to evaluate the 
commutator with the time ordered product and we use the similar commutation 
relation that holds for the local Wick products occurring in the sum. If this 
is carried out, then one finds that $[D(x_1, \dots, x_n), \varphi(y)] = 0$.  
Since the only elements of our
algebra $\cW(M, \g)$ that commute with all smeared field operators
$\varphi(f)$ are multiples of the identity \cite[Prop. 2.1]{hw}, 
we thus find that 
$D$ must in fact be given by a c-number distribution times the identity.
We define $t_{k_1\dots k_n}$ to be this c-number distribution. 
Now $t = t_{k_1 \dots k_n}$ can 
trivially be written as $t = \omega(D)$, for any Hadamard state, 
and each operator in the expression~\eqref{Ddef} for $D$ satisfies
\footnote{For the terms in the sum in eq.~\eqref{Ddef}, this 
follows from inductive hypothesis eq.~\eqref{twf} 
on the $t_{j_1 \dots j_n}$ with $|j| < |k|$, together with 
the fact that $\omega(\lno \varphi^{k_1}(x_1) \dots \varphi^{k_n}(x_n)
\rno_H)$ is smooth for all $k_1, \dots, k_n$.} T3. 
Hence, condition T3 holds also for $D$, thus showing that
$\WF(t_{k_1 \dots k_n}) \subset \cC_T(M, \g)$. We have therefore completed 
the induction step, thereby establishing that the Wick-expansion holds
for all multi-orders $(k_1, \dots, k_n)$, provided
only that T3 and T9 hold for the time ordered products. 

Conversely, if a
definition of time ordered products has been given that 
admits a Wick expansion of the form \eqref{wickexp1} with coefficients
satisfying \eqref{twf}, then
properties T3 and T9 will hold as well in the neighborhood of the total
diagonal on which the Wick expansion is defined.

Since the distribution $T^0$ defined on $M^{n+1} \setminus
\Delta_{n+1}$ by \eqref{T0} above satisfies properties T3 and T9
on $M^{n+1} \setminus \Delta_{n+1}$, it also admits a local Wick
expansion of the form \eqref{wickexp1}, i.e., on $\mathcal U_{n+1} \setminus
\Delta_{n+1}$ we have
\begin{equation} 
\label{wickexpT0}
T^0(\prod_{i=1}^{n+1} \varphi^{k_i}(x_i)) = 
\sum_{j_1 \dots j_{n+1}} {k \choose j}
t^0_{j_1 \dots j_{n+1}} (x_1, \dots, x_{n+1})
\lno \varphi^{k_1-j_1}(x_1) \dots \varphi^{k_{n+1}-j_{n+1}}(x_{n+1}) 
\rno_H.
\end{equation}
In the next subsection, we will reformulate the problem of extending
the algebra-valued distributions $T^0$ to the algebra-valued
distributions $T$ in terms of the extension of the c-number
distributions $t^0$ appearing in \eqref{wickexpT0} to the c-number
distributions $t$ appearing in \eqref{wickexp1}.

\subsection{Reformulation in terms of the extension of $t^0$} 

We return now to our inductive construction of time ordered products
We assume that all time ordered products involving $\leq n$ factors
have been constructed so as to satisfy our assumptions T1--T9 and we
consider an arbitrary time ordered product, $T$, in $(n+1)$
factors. As noted in subsection 3.1, property T8 will hold if and only
if $T$ is an extension to all of $M^{n+1}$ of the distribution $T^0$
on $M^{n+1} \setminus \Delta_{n+1}$ defined by \eqref{T0}. Since $T^0$
satisfies T1--T9 on $M^{n+1} \setminus \Delta_{n+1}$, we need only
check that our extension preserves these properties. As noted at the
end of subsection 3.1, we actually need only check that $T$ preserves
properties T1--T5 and T9, since T8 does not provide any additional
conditions on the extension and, by a suitable redefinition,
it is straightforward to ensure that T6 and T7 are satisfied.  
Furthermore, as shown in the previous
subsection, we may replace property T9 by the local Wick expansion
\eqref{wickexp1}. Thus, time ordered products satisfying all of our
conditions will exist if and only if the c-number distributions $t^0$
on $\mathcal U_{n+1} \setminus \Delta_{n+1}$ appearing in
\eqref{wickexpT0} can be extended to distributions $t$ on
$\mathcal U_{n+1}$ in such a way that the distribution $T$ defined by
\eqref{wickexp1} continues to satisfy properties T1--T5. It is
straightforward to check that this will be the case if and only if the
extensions $t$ satisfy the following 5 corresponding conditions:

\paragraph{t1 Locality/Covariance.}
The distributions $t$
are locally constructed from the metric in a covariant manner in the 
following sense. Let $\psi: N \to M$ be a causality-preserving
isometric embedding, and let $f$ be a test function supported in a 
sufficiently small neighborhood in the total diagonal of $N^{n+1}$. Then
we require that 
\begin{equation}
\label{psicond}
\psi^* t[\g_M](f) = t[\g_N](f), 
\end{equation}
where $\g_M$ and $\g_N$ are the metrics on $M$ and $N$ respectively, 
so that $\psi^* \g_M = \g_N$. 

\paragraph{t2 Scaling.}
The distributions $t = t_{j_1 \dots j_{n+1}}$ scale homogeneously
up to logarithmic terms,  in the sense that
there is an $N \in \mn$ such that 
\begin{equation}
\label{almost}
\lambda^{-d} t[\lambda^{-2} \g] 
= t[\g] + \sum_{h=1}^{N-1} \frac{\ln^h \lambda }{h!} \, v_h[\g], 
\end{equation}
where the $v_h$ are certain local and covariant distributions, 
and where $d = \sum j_i$. 

\paragraph{t3 Microlocal Spectral Condition.}
$\WF(t)\restriction_{\Delta_{n+1}} \perp T(\Delta_{n+1})$. 

\paragraph{t4 Smoothness.}
Let $\g^{(s)}$ be a smooth  family of  
metrics on $M$, depending smoothly on a parameter, and view
$t(s, x_1, \dots, x_{n+1}) = t[ \g^{(s)} ](x_1, \dots, x_{n+1})$
as a distribution on $\mr \times \mathcal{U}_{n+1}$. 
Then we require that 
\begin{equation}
\WF(t) \restriction_{\mr \times \Delta_{n+1}} 
\subset \{(s, \rho; x, k_1; \dots; x, k_{n+1}) \mid \sum k_i = 0, \,\,
\text{not all $k_i=0$}\}.
\end{equation}

\paragraph{t5 Analyticity.} 
If $\g^{(s)}$ is an analytic family of real analytic metric on
$\mathcal U_{n+1}$, then item t4 holds with the smooth wave front
set replaced by the analytic wave front set.

\medskip

\paragraph{Remarks.}

\noindent
(1) If we apply the differential operator 
$\lambda \partial_\lambda = \frac{\partial}{\partial \ln \lambda}$ 
a total of $N$ times to both sides of eq.~\eqref{almost}, then we obtain 
\begin{equation}
\label{scal}
\left(\lambda \partial_\lambda- d \right)^N
t[\lambda^{-2} \g] = 0.  
\end{equation}
Moreover, if we apply $\lambda \partial_\lambda=
\frac{\partial}{\partial \ln \lambda}$ only $h < N$ times to both
sides of \eqref{almost} and set $\lambda$ equal to one afterwards, the
we find that the local covariant distributions $v_h$ are given by
\begin{equation}
\label{lala}
v_h[\g] = (\lambda \partial_\lambda - d)^h 
\,t[\lambda^{-2}\g] |_{\lambda = 1}.
\end{equation}  
In fact, equation \eqref{scal}
is actually equivalent to \eqref{almost}, as one can 
see by rewriting the differential operator 
$\lambda \partial_\lambda - d$ as  
$\lambda^{d} \frac{\partial}{\partial \ln \lambda} \lambda^{-d}$ 
and then integrating \eqref{scal} $N$ times. 

\smallskip

\noindent
(2) In formulating conditions t3--t5, we have taken advantage
of the fact that on $\mathcal U_{n+1} \setminus \Delta_{n+1}$, we have
$t=t^0$, so $t$ is already known to satisfy the wave front set
conditions corresponding to T3--T5 on $\mathcal U_{n+1} \setminus
\Delta_{n+1}$. Consequently, we need only require $t$ to satisfy the
desired wave front set conditions on $\Delta_{n+1}$. Similarly,
conditions t1 and t2 also are already known to hold on $\mathcal
U_{n+1} \setminus \Delta_{n+1}$, so we need only check that $t$
satisfies these conditions in an arbitrarily small neighborhood of
$\Delta_{n+1}$.

\medskip

In summary, in this section we have reduced the problem of defining
time ordered products to the following question: Assume that time
ordered products involving $\leq n$ factors have been constructed so
as to satisfy our requirements T1--T9. Define $T^0$ by \eqref{T0}
and define the distributions $t^0$ on $\mathcal U_{n+1} \setminus
\Delta_{n+1}$ by \eqref{wickexpT0}. Can each $t^0$ be extended to
a distribution $t$ on $\mathcal U_{n+1}$ so as to satisfy requirements
t1--t5?

\section{Extension to the total diagonal}

Thus far, our analysis of time ordered products corresponds closely to
that given in \cite{bf}. The primary difference in our assumptions is
that we have imposed the requirement that time ordered products be
local, covariant fields (see T1) and that they satisfy certain
additional requirements concerning scaling behavior (see T2), and
smooth and analytic dependence on the metric (see T4 and T5). This has
resulted in some important differences in our analysis as compared
with \cite{bf}.  In particular, as a consequence of the
locality/covariance requirement, the Wick expansion of \cite{bf} in
terms of normal ordered products with respect to a quasifree Hadamard
state is not useful, so instead we introduced a local, covariant Wick
expansion in subsection 3.2. Nevertheless, all of the steps in the
analysis given in section 3 above are in close parallel with the
analysis of \cite{bf}.

As described at the end of section 3, our analysis will be completed
if we can extend the distributions $t^0$ to the total diagonal so that
they satisfy properties t1--t5. As is well known from quantum field
theory in Minkowski spacetime, straightforward attempts to extend
$t^0$ to the total diagonal give rise to formal expressions that do
not make sense as distributions. Therefore, one normally proceeds by
introducing some means of ``regularizing'' these formal expressions
and then extracting a well defined ``finite part'' (up to
renormalization ambiguities). In Minkowski spacetime, most approaches
to regularization/renormalization involve the use of Euclideanization
and/or momentum space methods, neither of which have a natural
generalization to (non-static) curved Lorentzian spacetimes. For this
reason, the authors of \cite{bf} employed the regularization procedure
of Epstein and Glaser, which is ``local'' in the sense that it uses
coordinate space methods that can be defined in a local region.

Nevertheless, the Epstein-Glaser method is not local in a strong
enough sense for our purposes, since we need to ensure that the
renormalized time ordered products will be local, covariant fields. A
key step in the Epstein-Glaser regularization procedure is the
introduction of certain ``cutoff functions'' of compact support in the
``relative coordinates'' that equal 1 in a neighborhood of the total
diagonal. Since the prescription for the extension of $t^0$ depends
upon the spacetime geometry throughout the region where the cutoff
functions are non-zero, the extension, $t$, at a point $p \in
\Delta_{n+1}$ will not depend only on the metric in an arbitrary small
neighborhood of $p$ and, thus, will not depend locally and covariantly
on the metric in the sense required by condition t1. There does not
appear to be any straightforward way of modifying the Epstein-Glaser
regularization procedure so that the resulting extension, $t$, will
satisfy property t1. In particular, serious convergence difficulties
arise if one attempts to shrink the support of the cutoff functions to
the total diagonal. In addition, the cohomological argument of
\cite{pr} also does not appear to admit a straightforward
generalization to curved spacetime.

Consequently, we shall proceed by a different route here. Our approach
to extend $t^0$ to the total diagonal is motivated by the idea
(essentially the ``equivalence principle'') that on sufficiently small
scales a curved space ``looks flat'', and that the divergences of
$t^0$ in curved spacetime should be of the same nature as the
corresponding $t^0$ in flat spacetime. However, this idea is not
correct as just stated because a curved space is not actually flat (no
matter on how small a scale one looks). Although it is true that the
leading order divergences of $t^0$ will be essentially the same as in
flat spacetime, in general there will be sub-leading-order divergences
that are sensitive to the presence of curvature and are different from
the divergences occurring for the corresponding $t^0$ in flat
spacetime. Nevertheless, we will see in subsection 4.1 below that any
local, covariant distributions that satisfies our scaling, smoothness,
and analyticity conditions admits a ``scaling expansion'' about the
total diagonal. This expansion expresses $t^0$ as a finite sum of
terms plus a remainder term with the properties that (i) each term in
the finite sum is a product of a curvature term times a distribution
in the relative coordinates that corresponds to a Lorentz invariant
distribution in Minkowski spacetime (which {\it can} be extended to
the total diagonal by Minkowski spacetime methods) and (ii) the
remainder term admits a unique, natural extension to the diagonal
by continuity. We
shall thereby obtain an extension of $t^0$ in subsection 4.2. In
subsection 4.3 we will then show that the resulting extension
satisfies all of the required properties t1--t5.

\subsection{The scaling expansion}

As indicated above, the key step that will enable us to extend $t^0$
is to perform a scaling expansion of it about the total diagonal. 
However, {\it a priori} it is not even clear what this means, since
$t^0$ is a distribution in $n+1$ variables, and it is not clear what
it would mean to perform any kind of ``expansion'' of a distribution
about the $4$-dimensional submanifold $\Delta_{n+1}$. 

The first step in obtaining the scaling expansion for $t^0$ is
to show that it is possible to fix one of its $n+1$ variables at a
value $x$, and view it as a distribution in the remaining $n$
variables, $y$, which play the same role as ``relative coordinates'' in
Minkowski spacetime. In other words, writing
\begin{equation}
x = x_1, \quad y = (x_2, \dots, x_{n+1}).
\label{y}
\end{equation}
we show that the (unextended) distribution $t^0$ possesses a
well-defined restriction to the submanifold
\begin{equation}
C_x = \{x\} \times (U^{n} \setminus (x, \dots x)), 
\end{equation}
where $U$ is a convex normal neighborhood of the point $x \in M$.
In Minkowski spacetime, this result would follow as an immediate
consequence of translation invariance. In our context, this result
follows from the microlocal spectral condition:
Since property T3 is known to hold for
$T^0$, it follows that the wave front set of $t^0$ is contained in the
set $\cC_T$. As can be seen from ``energy-momentum conservation
constraint'' in eq.~\eqref{gamtdef}, $\cC_T$
does not contain any elements of the form $(x, k; y, 0)$.  Since the
conormal bundle, $N^*C_x$ of the submanifold $C_x$ is spanned
precisely by such covectors, $\WF(t^0)$ does not have any elements in
common with $N^* C_x$. That the restriction exists is thus ensured by
\cite[Thm. 8.2.4]{h}. This restriction may be identified with a
distribution on $U^n \setminus (x, \dots, x)$.

We now shall obtain our scaling expansion of $t^0$. The basic idea is
to expand the $t^0[\g](x, \,\cdot\,)$ at a fixed point $x$ in terms of
the metric and its derivatives at $x$. The individual terms in the
so-obtained series will be seen to be given by sums of products of
local curvature terms at $x$ times Lorentz invariant distributions of
the relative coordinates.  The remainder for the suitably truncated
series will not have this simple form, but will turn out to be regular
enough to allow a unique extension.

To begin, we choose a convex normal neighborhood $U \subset M$ of $x$
and introduce Riemannian normal coordinates with respect to the metric
$\g$ around $x$. These coordinates are constructed by using the
exponential map to identify $U$ with a subset of the tangent space
$T_x M$, at $x$ and then identifying $T_x M$ with Minkowski
spacetime, $\mr^4$, by an isomorphism $e: T_x M \rightarrow \mr^4$. Thus,
the Riemannian normal coordinates of a point $\xi \in U$ are given by
\begin{equation}
\label{alphadef}
\alpha_x(\xi) = e \circ ({\rm exp}_x)^{-1}(\xi) \in \mr^4,  
\end{equation}
However, when it is not likely to cause confusion, we will slightly
abuse the notation by denoting all quantities---i.e., the point, its
Riemannian normal coordinates, and the corresponding vector in
Minkowski spacetime---simply by $\xi$. Similarly, the Riemannian
normal coordinates of $y = (x_2, \dots ,x_{n+1})$ (see eq.~(\ref{y})
above) will be denoted $\alpha_x(y)$, but when it is not likely to
cause confusion, we also will use $y$ to denote the Riemannian normal
coordinates of these points or the corresponding vector in
$\mr^{4n}$.

The choice of isomorphism $e: T_x M \rightarrow \mr^4$ is equivalent to
a choice of an orthonormal tetrad, $e^a_\mu$, at the point $x$.  Since
any other orthonormal tetrad is of the form $\Lambda_\mu^\nu e^a_\nu$
for some Lorentz transformation $\Lambda$, the Riemannian normal
coordinates, $\xi$, of a given point corresponding to the transformed
tetrad at $x$ are then given in terms of the original normal
coordinates by $\Lambda \xi$. Similarly, the Riemannian normal
coordinates, $y$, of a point in $U^n$ are obtained by Lorentz
transforming the coordinates of each point individually by $\Lambda$,
the result of which we shall denote as $\Lambda y$.

Now let $\g^{(s)}$ be the smooth 1-parameter 
family of metrics on $U$ whose coordinate components in 
Riemannian normal coordinates around $x$ are given by 
\begin{equation}
\label{gsdef}
g^{(s)}_{\mu\nu}(\xi) = g^{}_{\mu\nu}(s\xi). 
\end{equation}
Note that if $\chi_s$ is the map from $U$ into itself given by 
$\xi \to s\xi$ in Riemannian normal 
coordinates about $x$, then this family of metrics 
can be alternatively written as 
\begin{equation}
\g^{(s)} = s^{-2} \chi_s^* \g.
\end{equation}
Note also that the definition of the above family of metrics does not
depend on any additional data besides the specification of the point 
$x$ and the metric itself. In particular, it does not 
depend on our choice of tetrad at $x$.

By a slight generalization of the microlocal argument given at the
beginning of this subsection, it follows from the fact that $T^0$
satisfies properties T3 and T4 on $U^{n+1} \setminus \Delta_{n+1}$
that $t^0[ \g^{(s)} ](x, \,\cdot\,)$ makes sense as a family of
distributions on $U^n \setminus (x, \dots, x)$ that is parametrized by
$(s,x)$. Furthermore, when smeared with a test function, $f$, in $y$,
it follows that $t^0[ \g^{(s)} ](x,f)$ is smooth in $(s,x)$.  In
addition, since differentiation does not increase the size of the
wave front set, derivatives of $t^0$ with respect to $s$ also make
sense as distributions on $U^n \setminus (x, \dots, x)$ that are
parametrized by $(s,x)$. Hence, for any $k$ and any arbitrary, but
fixed $x \in M$, we can define a distribution on $U^n \setminus (x,
\dots, x)$ by
\begin{equation}
\label{****}
\tau^0_k[\g](x, \,\cdot\,) =
\frac{d^k}{ds^k}
t^0\left[ \g^{(s)} \right] (x, \,\cdot\,)
\Big|_{s=0}. 
\end{equation}
It follows that for any given natural number $m \ge 0$, we have
the following Taylor expansion with remainder:
\begin{equation}
\label{**}
t^0 = \sum_{k=0}^m \frac{1}{k!}\,\tau^0_k + r^0_m, 
\end{equation}
where
\begin{equation}
\label{remain} 
r^0_m[\g](x, \,\cdot\,) = \frac{1}{m!} 
\int_0^1 (1-s)^{m}
\frac{d^{m+1}}{ds^{m+1}}
t^0\left[\g^{(s)}\right](x, \,\cdot\,)
\, ds.   
\end{equation}
Formula \eqref{**} is actually our desired scaling expansion of
$t^0$. However, as it stands, \eqref{**} is merely an identity that
would hold for any distribution in the variables $(s,x,y)$ that
satisfies suitable wave front set conditions. The important properties
of this formula for our distributions $t^0$ are stated in the
following theorem.

\begin{thm}
\label{1.4}
\begin{enumerate}
\item[(i)] $\tau^0_k(x, \,\cdot\,)$ and $r^0_m(x, \,\cdot\,)$ are
distributions on $U^n \setminus (x, \dots, x)$ which are locally
constructed from the metric in a covariant way in the sense that 
eq.~\eqref{psicond} holds for all diffeomorphisms that leave the point $x$ 
invariant.
\item[(ii)] 
We have the decomposition
\begin{equation}
\label{taudec}
\tau^0_k (x, y) = \sum
C(x) \cdot \alpha^{*}_x u^0(y), 
\end{equation}
where the sum is finite. Here, $C \equiv C^{\mu_1 \dots \mu_l}$ denote the 
coordinate components of certain curvature tensors 
in Riemannian normal coordinates about $x$ and 
the $u^0 \equiv u^0_{\mu_1 \dots \mu_l}$ are 
Lorentz-invariant tensor
valued distributions defined on 
$\mr^{4n}$ with the origin removed, that is,  
\begin{equation}
u^0_{\mu_1 \dots \mu_l}
(\Lambda \,\cdot\,)
= \Lambda_{\mu_1}^{\nu_1} \dots \Lambda_{\mu_l}^{\nu_l}
u^0_{\nu_1 \dots \nu_l}
(\,\cdot\,)
\end{equation}
for any Lorentz-transformation $\Lambda$. 
The local curvature tensors $C$ arise as a sum of monomials 
in $g_{ab}, R_{abcd}, \dots, \nabla_{(e_1} 
\dots \nabla_{e_{k-2})} R_{abcd}$. In 
the case considered here with no dimensional parameters, each 
monomial contains precisely $k$ coordinate derivatives of the metric. 
\item[(iii)]
$\tau^0_k$ and $r^0_m$ scale almost homogeneously
under rescalings of the {\it metric} with degree $d$.
\item[(iv)]
The distributions
$u^0$ scale almost homogeneously with degree $d-k$ under 
{\it coordinate} rescalings in the sense that
there exists an $N \in \mn$ such that 
\begin{equation}
\label{***}
S_{d - k}^N u^0 = 0,  
\end{equation}
where $S_\rho^N = (\sum \xi_i^\mu \partial/\partial \xi_i^\mu + \rho)^N$.
\item[(v)] The scaling degree of $r^0_m(x, \,\cdot\,)$ is less or
equal than $d - m - 1$, i.e., the distributions
$\lambda^{d-m-1+\delta}r^0_m(x, \lambda\,\cdot\,)$, viewed as
distributions on $\mr^{4n} \setminus 0$ via the pull-back by
$\alpha_x$, tend to the zero distribution as $\lambda \searrow 0$ for
all $x$ and all $\delta > 0$.
\end{enumerate}
\end{thm}

\paragraph{Remark.}
We note that (iv) means that 
$u^0$ scales homogeneously 
with degree $d-k$ under
a rescaling of the coordinates, up to logarithmic terms. Namely, 
simple integration of \eqref{***} gives that 
\begin{equation}
\lambda^{d-k} u^0(\lambda \, \cdot\,)
= u^0(\, \cdot \,)
+ \sum_{h = 1}^{N-1} 
\frac{\ln^h \lambda}{h!} \,S_{d-k}^h u^0 (\, \cdot\,). 
\end{equation}  
This implies in particular that the scaling degree of 
$u^0$ at the origin is $d - k$. 

\begin{proof}
Item (i) follows directly from the fact that $t^0$ is local and
covariant on its domain of definition.  

To prove (ii), we consider, first, the case where all the components
of $\g$ in our Riemannian normal coordinates are polynomials in the
Riemannian normal coordinates $\xi$ in a neighborhood of $x$. Then we
may characterize $\g$ by its components $g_{\mu\nu}$ at $x$ together
with the components of the coordinate derivatives, $g_{\mu\nu,
\sigma_1 \sigma_2 \dots}$, at $x$, only finitely many of which are
nonzero. We may thus view $t^0$ as being a function of these
quantities, and we express this by writing
\begin{equation}
t^0[\g](x, \,\cdot\,) = t^0[g_{\mu\nu}, \dots, 
g_{\mu\nu, \sigma_1\sigma_2 \dots \sigma_l}, \dots] (x, \,\cdot\,), 
\end{equation}
Now smear with a test function, $f$, on $U^n \setminus (x, \dots, x)$.
Since $t^0[\g](x,f)$ depends smoothly on the metric---and hence is a
smooth function of the finite number of variables 
$$g_{\mu\nu}(x), \dots, g_{\mu\nu, \sigma_1 \sigma_2 \dots \sigma_l}(x), 
\dots$$ 
on which it depends---we obtain
\begin{eqnarray}
\label{monster}
\tau^0_k[\g](x,f) 
&=& \partial_s^k
t^0\left[ \g^{(s)}\right](x,f) \Big|_{s=0}\nonumber\\
&=& \partial_s^k
t^0[g_{\mu\nu}, \dots, 
s^l g_{\mu\nu, \sigma_1\sigma_2 \dots \sigma_l}, \dots](x,f) 
\Big|_{s=0}
\nonumber\\
&=&  k! \sum_{l_1 + 2l_2 + \dots ml_m = k} 
\frac{\partial^{l_1 + \dots +l_m}t^0[\dots](x,f)}{
\partial^{l_1} g_{\mu\nu, \sigma_1} 
\dots 
\partial^{l_m} g_{\mu\nu, \sigma_1\sigma_2 \dots \sigma_m}} 
\prod_j [g_{\mu\nu,\sigma_1 \sigma_2 \dots \sigma_j}(x)]^{l_j},
\end{eqnarray}
where $[\dots]$ stands for $[g_{\mu\nu}, 0, 0, \dots]$. 

We may rewrite this equation as
\begin{equation}
\label{unicdef}
\tau^0_k(x,y)
=\sum C(x) \cdot  
\alpha_x^{*} u^0(y),  
\end{equation}
where the $C \equiv C^{\mu_1 \dots \mu_l}$ are monomials in
$g_{\mu\nu,\sigma_1 \sigma_2 \dots \sigma_m}$, which have the property
that the total number of derivatives of $g_{\mu\nu}$ appearing in each $C$
is equal to $k$, and where ${u}^0 \equiv u^0_{\mu_1 \dots
\mu_l}$ are tensor-valued distributions on $\mr^{4n}$ minus the
origin\footnote{Actually, in the above construction the distributions
$u^0$ are automatically defined only in the neighborhood of the origin
in $\mr^{4n}$ (minus the origin itself) corresponding to the
neighborhood, $U \subset M$ on which the Riemannian normal coordinates
are defined. However, by modifying $\g$ outside of a neighborhood of
the origin if necessary, we may assume without loss of generality that
the Riemannian normal coordinates are globally defined and that $u^0$
is defined everywhere on $\mr^{4n} \setminus 0$.}, which are
independent of $\g$.  Since we are working in Riemannian normal
coordinates, the $m$-th coordinate derivates $g_{\mu\nu, \sigma_1
\sigma_2 \dots \sigma_m}$ of the metric tensor at $x$ can be rewritten
as the coordinate components at $x$ of a local curvature term that is
polynomially constructed from the metric, the curvature tensor and its
derivatives at $x$. Moreover, such a curvature term must involve
precisely $m$ derivatives of the metric.  Hence, by our formula
\eqref{monster}, we conclude that $C^{\mu_1 \dots \mu_l}$ corresponds
to a curvature term $C^{a_1 \dots a_l}$ which arises as a sum of
monomials in $R_{abcd}, \dots, \nabla_{(e_1} \dots \nabla_{e_{k-2})}
R_{abcd}$, each of which contains precisely $k$ derivatives of the
metric.

We would next like 
to show that the distributions ${u}^0$
are Lorentz invariant. For this, we consider the diffeomorphism
$\psi_\Lambda$ on $U \subset M$ given by 
$\xi \to \Lambda \xi $ where $\Lambda$ is a Lorentz transformation, 
and where the point $\xi \in U$ has been identified with its 
Riemannian normal coordinates about $x$. 
By definition, this diffeomorphism will leave the point $x$ 
invariant, so we may apply item (i) to this diffeomorphism. From 
this, we get that
\begin{equation}
\sum C^{\mu_1 \dots \mu_l}(x)
{u}^0_{\mu_1 \dots \mu_l}(\Lambda \,\cdot\,)
= 
\sum C^{\mu_1 \dots \mu_l}(x)\Lambda^{\nu_1}_{\mu_1}
\dots \Lambda^{\nu_l}_{\mu_l}
{u}^0_{\nu_1 \dots \nu_l}(\,\cdot\,).   
\end{equation}
Since this holds for all metrics, this means that ${u}^0$ must be
Lorentz invariant. This proves (ii) for all metrics whose components
in Riemannian normal coordinates are polynomials in the Riemannian normal
coordinates $\xi$ in a neighborhood of $x$.

Now consider an arbitrary smooth metric $\g$. In a compact
neighborhood, $K$, of $x$, we can, for each $n$, find a
metric ${\bf q}^{(n)}$ that is polynomial in $\xi$ and is such that
everywhere within $K$ we have $|g_{\mu\nu, \sigma_1 \sigma_2
\dots \sigma_m} - q_{\mu\nu, \sigma_1 \sigma_2 \dots \sigma_m}^{(n)}| 
< 2^{-n}$ for all $m \leq n$. Let $\psi:\mr \rightarrow [0,1]$ be a smooth
function with support in $[-1,1]$ satisfying $\psi(-v) =
\psi(v)$ and also satisfying $1 - \psi(v) = \psi(1-v)$ for all $v \in
[0,1]$. Set ${\bf h}^{(0)} = \g$ and for $s \neq 0$ but in a sufficiently
small neighborhood of $0$ define
\begin{equation}
{\bf h}^{(s)} = \sum_n \psi(|1/s| - n) {\bf q}^{(n)} 
\end{equation}
(Note that at each $s$, there can be at most two terms in this sum
which are nonvanishing.) Then it is straightforward to show that ${\bf
h}^{(s)}$ is a one parameter family of smooth metrics that depends
smoothly on $s$. Consequently, each $\tau^0_k[{\bf h}^{(s)}](x,
\,\cdot\,)$ varies smoothly with $s$. However, we have already proven
that eq.~(\ref{taudec}) holds for all $s \neq 0$. By the smoothness
property t3 applied to $t^0$, it follows that eq.~(\ref{taudec})
continues to hold at $s = 0$, thus proving property (ii) for an
arbitrary smooth metric $\g$.

Property (iii) is a direct consequence of the fact that $t^0$
satisfies the scaling property t2. To see this, we note that
\begin{equation}
\left(\lambda\partial_\lambda - d
\right)^N 
\tau^0_k[\lambda^{-2} \g] 
= \partial^k_s 
\left(\lambda \partial_\lambda- d
\right)^N 
t^0 \left[ \lambda^{-2} \g^{(s)} \right]
\big|_{s=0} = 0, 
\label{tauscale}
\end{equation}
since $t^0$ satisfies t2. This establishes (iii) for $\tau^0_k$. 
That $r^0_m$ satisfies (iii) then follows immediately from eq.~(\ref{**}).

To prove (iv), we note that (i) implies that
\begin{eqnarray}
\label{ppp}
\tau^0_k[\lambda^{-2} \g]  &=& \partial^k_s
t^0 \left[\lambda^{-2} \g^{(s)} \right] \big|_{s=0}
= \partial^k_s
t^0 \left[(\lambda s)^{-2} \chi_s^* \g \right] \big|_{s=0}\nonumber\\
&=& 
\chi_{\lambda^{-1}}^* \partial^k_s
t^0\left[\g^{(\lambda s)}\right] \big|_{s=0} = \lambda^k 
\chi_{\lambda^{-1}}^* \tau^0_k[\g] 
\end{eqnarray}
By eq.~\eqref{tauscale}, 
the differential operator $(\lambda \partial_\lambda - d)^N$ 
annihilates the left side of eq.~\eqref{ppp}. This implies that
\begin{equation}
\label{sc1}
0 = \left(\lambda\partial_\lambda- d\right)^N 
\lambda^k 
\chi_{\lambda^{-1}}^* \tau^0_k[\g]
= \lambda^k \left(\lambda \partial_\lambda- d + k \right)^N 
\chi_{\lambda^{-1}}^* \tau^0_k[\g].
\end{equation}
Substituting the decomposition of $\tau^0_k$ into the expression on 
the right side, we obtain
\begin{eqnarray}
\label{sc2}
0 &=& 
\sum C(x) \cdot
\left(\lambda \partial_\lambda- d + k\right)^N 
{u}^0(\lambda \,\cdot\,) \nonumber \\
&=& 
\sum C(x) \cdot S_{d-k}^N
{u}^0(\lambda \,\cdot\,).
\end{eqnarray}
Since this holds for arbitrary metrics $\g$, it follows that that
$S_{d-k}^N {u}^0 = 0$, as we desired to show.

In order to establish the estimate on the scaling degree 
for $r^0_m$, item (v), we first use 
eq.~\eqref{remain} along with the
same arguments as in (iv) to write  
\begin{equation}
r^0_m(x, \lambda \,\cdot\,)
= \frac{\lambda^{m+1}}{m!}\int_0^1 (1 - \mu)^{m}
\partial^{m+1}_s
t^0\left[\lambda^2 \g^{(s)} 
\right](x, \,\cdot\,) \big|_{s=\lambda\mu}
\, d\mu. 
\end{equation}
Using the fact that $t^0$ satisfies property t2, we have
(see eqs.~\eqref{almost} and~\eqref{lala})
\begin{equation}
\label{psidef}
r^0_m(x, \lambda \,\cdot\,) = 
\lambda^{m + 1 - d} \sum_{l=0}^{N-1}
\ln^l \lambda \, \psi^0_l(\lambda, x, \,\cdot\,), 
\end{equation} 
where 
\begin{equation}
\psi^0_l(\lambda, x, \,\cdot\,) \mydef \,\, \frac{1}{l!m!}
\int_0^1 (1 - \mu)^{m}
\partial^{m+1}_s (v \partial_v - d)^l\,
t^0\left[ v^2\g^{(s)} \right](x, \,\cdot\,) \big|_{s =\lambda \mu, v=1}
d\mu.
\label{psidef2}
\end{equation}
If
$f$ is a smooth test function on $U^n$ whose support does not contain
the point $(x, \dots, x)$, then by wave front set arguments similar to
those given above, it follows from the fact that $T^0$ satisfies
conditions T3 and T4 that the quantities $\psi^0_l (\lambda,
x, f)$ are smooth in $\lambda$ in a neighborhood of zero. This
immmediately implies (v).
\end{proof}

\paragraph{Remarks.}

(1) As stated in property (v) of the above theorem, if we carry the
scaling expansion, eq.~(\ref{**}), to higher order (i.e., larger $m$),
the remainder term $r^0_m$ will have a lower scaling degree. However,
it should be noted that the wave front set of $t^0$ is determined by
the null geodesics of the curved spacetime metric $\g$ whereas the
wave front set of each $\tau_k$ is similarly determined by the null
geodesics of the flat spacetime metric associated with the exponential
map at $x$. Since the null geodesics of these two metrics do not, in
general, coincide (with the exception of the null geodesics passing
through $x$ itself), it is clear that $r^0_m$ remains fundamentally
distributional in nature no matter how large $m$ is chosen. It also
should be noted that it is {\it not} claimed in Thm. \ref{1.4} that
$r^0_m$ converges to zero in any sense (even for an analytic
spacetime) as $m \rightarrow \infty$. Thus, eq.~(\ref{**}) should be
viewed only as a ``scaling expansion'' with the properties specified
in Thm. \ref{1.4}, not as a convergent power series.

\smallskip

\noindent
(2) If we combine eqs.~(\ref{**}) and (\ref{taudec}), we obtain an expansion
of $t^0$ of the general form
\begin{equation}
\label{===}
t^0(x,y) = \sum  C(x) \cdot \alpha_x^* u^0(y) + r^0_m(x,y).
\end{equation}
If the terms in the sum in \eqref{===} are ordered by the engineering
dimension of the curvature terms, $C$, then the first term in the
expansion has $C=1$ and the corresponding distribution $u^0$ is the
``scaling limit'' at $x$ of the distributions $t^0$ in the sense of
Fredenhagen and Haag \cite{fh}.  The higher order terms in the
expansion then give corrections to the the scaling limit, organized in
powers of the curvature tensor and its derivatives.  If dimensionful
parameters are present in the theory, then the scaling expansion will
be organized in terms of products of powers of the curvature and the
dimensionful parameters. Our scaling expansion is also closely related
to the ``momentum space representation'' of the Feynman propagator
and its powers (see remark (3) below)
given in \cite{pf}, since the Lorentz invariant distributions, $u^0$,
on Minkowski spacetime occurring in our expansion can be given a
momentum space representation.

\smallskip

\noindent
(3) For the Feynman propagator and its powers, the scaling 
expansion can be explicitly calculated from known properties
of the Hadamard expansion. We will illustrate this with 
two examples. The first
example is the simplest nontrivial time ordered product,
$T^0(\varphi(x) \varphi(y))$.  Its Wick-expansion is given by
\begin{equation}
\label{HF}
T^0(\varphi(x) \varphi(y)) = \,\, \lno \varphi(x) \varphi(y) \rno_H
+ H_F(x, y) \myid, 
\end{equation}  
where $H_F = H - \i \Delta^{\rm adv}$ is the ``local Feynman
parametrix'', where $H$ is the Hadamard parametrix, eq.~(\ref{hadamp}),
and $\Delta^{\rm adv}$ is the advanced Green's function. Thus, the
only nontrivial distribution $t^0$ occurring in this expansion is
\begin{equation}
t^0(x,y) = H_F(x, y) = U(x, y)(\sigma + \i 0)^{-1}
+ V(x, y) \ln (\sigma + \i 0),    
\end{equation}
where $U$ and $V$ are as in the Hadamard parametrix (see
eq.~\eqref{hadamp}). The first few terms, $\tau^0_k$, in the scaling
expansion for $t^0 = H_F$ are easily found from the expansions for $U$
and $V$ given in \cite{deWitt} and many other references. Modulo an
overall constant, one finds
\begin{eqnarray*}
\tau^0_0(x, y) &=&  
(\eta_{\mu\nu}\xi^\mu \xi^\nu + 
\i 0)^{-1} \\
\tau^0_1(x, y) &=& 0\\
\tau^0_2(x, y) &=& 
\tfrac{1}{12}R^{\sigma\rho}(x)\xi_\sigma
\xi_\rho(\eta_{\mu \nu} \xi^\mu \xi^\nu + \i 0)^{-1}
-\tfrac{1}{24}R(x) \ln(\eta_{\mu \nu} \xi^\mu \xi^\nu + \i 0),    
\end{eqnarray*}
where, as above, $\xi^\mu$ denotes the Riemannian normal coordinates of
$y$ relative to $x$. Thus, in this example, our scaling expansion
corresponds to the usual short distance approximation to the singular
part of the Feynman propagator (see, e.g., \cite{chr}).

Our second example is the time ordered product $T^0(\varphi^2(x) 
\varphi^2(y))$. Its Wick-expansion is given by 
\begin{equation}
T^0(\varphi^2(x) \varphi^2(y)) = \,\, \lno \varphi^2(x) \varphi^2(y) \rno_H
+ 2H_F(x, y) \lno \varphi(x) \varphi(y) \rno_H + H_F(x, y)^2 \myid. 
\end{equation}
The only new $t^0$ arising in this expansion is the ``fish graph'',
$t^0 = H_F^2$, a solution to the renormalization of which was
found by B.~S.~Kay~\cite{k} prior to the commencement of the present work
and played a role in the development of the present work.
As can be seen from the above expansion for $H_F$, the
first few coefficients, $\tau_k^0$, for the fish graph are, modulo an
overall constant,
\begin{eqnarray*}
\tau^0_0(x, y) &=&  
(\eta_{\mu\nu}\xi^\mu \xi^\nu + 
\i 0)^{-2} \\
\tau^0_1(x, y) &=& 0\\
\tau^0_2(x, y) &=& 
\tfrac{1}{6} R^{\sigma\rho}(x)\xi_\sigma
\xi_\rho(\eta_{\mu \nu} \xi^\mu \xi^\nu + \i 0)^{-2}
-\tfrac{1}{12} R(x) 
(\eta_{\mu \nu} \xi^\mu \xi^\nu + \i 0)^{-1}
\ln(\eta_{\sigma \rho} \xi^\sigma \xi^\rho + \i 0).
\end{eqnarray*}

It is easily seen that in both examples, the distributions $\tau^0_k$
are local, covariant distributions of the form claimed in (ii)---i.e.,
they are sums of terms of the form $C(x) \cdot \alpha_x^*u^0(y)$ with
$u^0$ a Lorentz-invariant Minkowski space distribution---and satisfy
the scaling properties specified in Thm. \ref{1.4}.

\smallskip
\noindent
(4) The above scaling expansion was carried out for the scalar
distributions $t^0$. It is straightforward to check that it also holds
for the extended distributions $t$ that will be defined in the next
subsection.  Much more generally, it should be possible to perform a
similar scaling expansion for arbitrary local covariant fields that
satisfy appropriate wave front set properties. This should yield a
generalized operator product expansion in curved spacetime. We are
currently investigating the properties of such an expansion.

\subsection{Extension of $t^0[\g]$}

Theorem 4.1 of the previous subsection provides the necessary
machinery to achieve our goal of extending $t^0$ in such a way that
properties t1--t5 are satisfied. The basic idea is simply to suitably
extend each term in the scaling expansion, eq.~\eqref{**}. Each
$\tau^0_k$ in that equation is of the form (\ref{taudec}) and hence
can be extended to the total diagonal by extending the Minkowski
spacetime distributions $u^0$ to the origin. This can be achieved by
standard methods used in Minkowski spacetime. On the other hand, if
$m$ is chosen sufficiently large, the remainder term $r^0_m$ will have
sufficiently low scaling degree that it can be extended to the total
diagonal by continuity. The proof that the so-obtained extension $t$
satisfies properties t1--t5 will be given in the next subsection.

The key result needed to extend each $\tau^0_k$ is the following:

\begin{lemma}\label{help}
Let $u^0 \equiv u^0_{\mu_1 \dots \mu_l}(y)$ with $y = (\xi_1, \dots,
\xi_n)$ be a Lorentz invariant tensor-valued distribution on $\mr^{4n}
\setminus 0$ which scales almost homogeneously with degree $\rho$
under coordinate rescalings, i.e.,
\begin{equation}
S_{\rho}^N u^0 = 0 \quad \text{ 
for some natural number $N$.} 
\end{equation}
where $S_\rho^N = (\sum \xi_i^\mu \partial/\partial \xi_i^\mu + \rho)^N$. Then
$u^0$ has a Lorentz invariant extension, $u$, to a distribution on
$\mr^{4n}$ which also scales almost homogeneously with degree $\rho$
under rescalings of the coordinates.
\end{lemma}

\begin{proof}
We will first extend $u^0$ using the Epstein-Glaser prescription.
This extension need not satisfy either the scaling or Lorentz
invariance properties. However, we will show that the extension can be
modified, if necessary, so as to scale almost
homogeneously\footnote{For distributions with an {\it exactly}
homogeneous scaling, this result has previously been obtained in
\cite[Thms. 3.2.3 and 3.2.4]{h}. Thus, our theorem generalizes this
result to the case of almost homogeneous scaling.} with degree
$\rho$. We will then show that the resulting extension can be further
modified, if necessary, so as to be Lorentz invariant while retaining
the almost homogeneous scaling with degree $\rho$.

Choose an arbitrary smooth function $w$ of compact support
on $\mr^{4n}$ which is equal to one in a neighborhood of the origin. 
For any test function $f \in \cD(\mr^{4n})$ we set 
\begin{equation}
(Wf)(y) = f(y) - w(y) \sum_{|\alpha| \le \rho-4n} 
y^\alpha \partial_\alpha f(0)/\alpha!,  
\end{equation}
where we use the usual multi-index notation. It follows from $S_\rho^N
u^0 = 0$ that $u^0$ has scaling degree $\rho$, so by
\cite[Thm. 5.3]{bf}, we can define an extension, $u$, of $u^0$ to
$\mr^{4n}$ by setting
\begin{equation}
u(f) = u^0(Wf).
\end{equation}
It follows that the scaling degree of $u$ is $\rho$ \cite[Thm. 5.3]{bf}, but
it need not hold that $u$ scales almost homogeneously with degree
$\rho$, i.e., there is no guarantee that $S_\rho^M u = 0$ for some
natural number $M$. However, one can calculate that
\begin{equation}
WS_\rho^N f(y) - S_\rho^N W f(y) = 
\sum_{|\alpha| \le \rho-4n} \psi^\alpha(y) \partial_\alpha f(0)
\end{equation}
for some smooth functions $\psi^\alpha$ whose support does not 
contain the origin. From this it follows immediately that
\begin{equation}
S_\rho^N u = 
\sum_{|\alpha| \le \rho-4n} c^\alpha \partial_\alpha\delta,
\label{uscale}
\end{equation}
where $c^\alpha = (-1)^{|\alpha|} u^0(\psi^\alpha)$. 

We now define a modified distribution $u'$ by  
\begin{equation}
u' = u - \sum_{|\alpha| \le \rho - 4n - 1} \frac{c^\alpha}{(\rho - 4n -
|\alpha|)^N} \partial_\alpha \delta.
\end{equation} 
Using the fact
that $S_\rho \partial_\alpha \delta = (\rho - 4n - |\alpha|) \partial_\alpha
\delta$, we find  
\begin{equation}
S_\rho^N u' = \sum_{|\alpha| = \rho - 4n} 
c^\alpha \partial_\alpha \delta.
\end{equation}
If we apply the operator $S_\rho$ to both
sides of the above equation, then we 
get that $S_\rho^{N+1} u' = 0$, because
\begin{equation}
S_\rho \partial_\alpha \delta = 0
\quad \text{for $|\alpha| = \rho-4n$.}
\end{equation}
This means that $u'$ is an extension of $u^0$ with the desired almost
homogeneous scaling. For notational simplicity, we will drop the
``prime'' in the following and denote this modified extension as $u$.

We now investigate the Lorentz transformation properties of
$u$. Restoring the tensor indices on $u$, we find by a calculation
similar to eq.~(\ref{uscale}) above that for any test function $f \in
\cD(\mr^{4n})$ and any Lorentz transformation, $\Lambda$, we have
\begin{equation}
\label{71}
{u}_{\mu_1 \dots \mu_l}(f) - 
\Lambda^{\nu_1}_{\mu_1} \dots \Lambda^{\nu_l}_{\mu_l}
{u}_{\nu_1 \dots \nu_l}(R(\Lambda)f) 
= \sum_{|\alpha| \le \rho- 4n} b^\alpha_{\mu_1 \dots \mu_l}(\Lambda) 
\partial_\alpha 
\delta(f),  
\end{equation}
where $(R(\Lambda)f)(y) = f(\Lambda y)$ and the $b^\alpha_{\mu_1 \dots
\mu_l}(\Lambda)$ are complex constants, which would vanish if and only
if the distribution $u$ were Lorentz invariant. We now apply the
differential operator $S_{\rho}^{N+1}$ to both sides of the above
equation. Since $S_{\rho}$ is itself a Lorentz invariant operator, we
have $R(\Lambda) S_{\rho} = S_{\rho} R(\Lambda)$. Therefore, since $S_{\rho}^{N+1}
{u} = 0$, the operator $S_\rho^{N+1}$ annihilates the left side of eq.~\eqref{71},
so we obtain
\begin{equation}
0 = S_{\rho}^{N+1} 
\sum_{|\alpha| \le \rho- 4n} b^\alpha_{\mu_1 \dots \mu_l}(\Lambda) 
\partial_\alpha \delta = 
\sum_{|\alpha| \le \rho- 4n} (\rho- 4n - |\alpha|)^{N+1}
b^\alpha_{\mu_1 \dots \mu_l}(\Lambda) 
\partial_\alpha \delta.  
\end{equation}
It follows immediately that $b^\alpha_{\mu_1 \dots \mu_l}(\Lambda) 
= 0$, except possibly when $|\alpha| = \rho- 4n$. Thus, we have
\begin{equation}
{u}_{\mu_1 \dots \mu_l}(f) - 
\Lambda^{\nu_1}_{\mu_1} \dots \Lambda^{\nu_l}_{\mu_l}
{u}_{\nu_1 \dots \nu_l}(R(\Lambda)f) 
= b_{\mu_1 \dots \mu_l}^{\nu_1 \dots \nu_{\rho-4n}}(\Lambda) 
\partial_{\nu_1} \dots \partial_{\nu_{\rho-4n}}
\delta(f)  
\end{equation}
for all $f$ and all Lorentz-transformations $\Lambda$. Using this equation, 
one finds the following transformation property for $b(\Lambda)$,
\begin{equation}
b(\Lambda_1 \Lambda_2) = b(\Lambda_1) + D(\Lambda_1)b(\Lambda_2), 
\end{equation}
where we have now dropped the tensor-indices and where $D$ denotes the
tensor representation of the Lorentz-group on $(\otimes^l \mr^4)^*
\otimes (\otimes^{\rho-4n} \mr^4)$.  It then follows by the
cohomological argument given in \cite{pr} that this relation implies
that $b$ can be written in the form
\begin{equation}
b(\Lambda) = a - D(\Lambda)a \quad \forall \Lambda, 
\end{equation}
where $a$ is an element in $(\otimes^l \mr^4)^* \otimes  
(\otimes^{\rho-4n} \mr^4)$, not depending on $\Lambda$. 
This enables us to define the modified extension
\begin{equation}
{u}_{\mu_1 \dots \mu_l}' = 
{u}_{\mu_1 \dots \mu_l} - 
a_{\mu_1 \dots \mu_l}^{\nu_1 \dots \nu_{\rho-4n}} 
\partial_{\nu_1} \dots \partial_{\nu_{\rho-4n}}
\delta,   
\end{equation}
where we have now restored the tensor indices. It is easily checked
that $u'$ is Lorentz invariant and satisfies $S_{\rho}^{N+1} {u}' = 0$.
We have therefore accomplished the goal of constructing the desired
extension of ${u}^0$.

\end{proof}

Some analyticity properties of $u$ and its Fourier transform that 
follow from its scaling behaviour are established in Appendix B. These
results, however, will not be needed in our present analysis.

We now can give our prescription for extending $t^0$. Let $d$ denote
the scaling degree of $t^0$, let $m = d - 4n$, and consider the
expansion eq.~(\ref{**}). By theorem 4.1, each $\tau^0_k$ appearing in
this expansion takes the form
\begin{equation}
\tau^0_k(x, y) = 
\sum C(x) \cdot \alpha_x^{*} u^0(y)  
\end{equation}
where the sum is finite. We extend $\tau^0_k$ to a distribution
$\tau_k$ on $U^n$ by choosing an extension, $u$, of each $u^0$ that
satisfies the properties of Lemma \ref{help} and defining
\begin{equation}
\label{expan}
\tau_k(x, y) \equiv 
\sum C(x) \cdot \alpha_x^{*} {u}(y).
\end{equation}
Although $\tau_k$ has been constructed as a distribution in $y$ that
is parametrized by $x$, it is straightforward to check that $\tau_k$
may also be viewed as a distribution jointly in $x$ and $y$.

On the other hand, we know by property (v) of Theorem 4.1 that the
scaling degree of $r^0_m$ is less or equal to $4n - 1$. Therefore we
can apply \cite[Thm. 5.2]{bf} to conclude that $r^0_m(x, \,\cdot\,)$ has
a unique extension, $r_m(x, \,\cdot\,)$ to all of $U^n$ with the same
scaling degree for any given point $x$. 
This extention is given by
\begin{equation}
\label{seq}
r_m(f) = \lim_{j\to \infty} r^0_m(\vartheta^{(j)} f), 
\end{equation} 
where $\vartheta^{(j)}$ is a sequence of smooth functions with support
in $U^{n+1} \setminus \Delta_{n+1}$, which are identically one outside
neighborhoods ${\cal U}_{n+1}^{(j)}$ of $\Delta_{n+1}$, with ${\cal
U}_{n+1}^{(j)}$ shrinking to $\Delta_{n+1}$ as $j$ goes to
infinity. By the scaling properties of $r^0_m$, this limit exists in
the weak sense, and is independent of the particular choice of cutoff
functions $\vartheta^{(j)}$ (see \cite[Thm. 5.2]{bf}). Again, it can
be shown that this extension defines a distribution jointly in $x$ and
$y$.

Our extension, $t$, is then defined by
\begin{equation}
t = \sum_{k=0}^m \frac{1}{k!}\,\tau_k + r_m.
\label{tdef}
\end{equation}
Our remaining task is to show that $t$ satisfies properties t1--t5.

\subsection{Proof that $t$ satisfies properties t1--t5}

As we now shall show, it is relatively straightforward to prove 
that the extension, $t$, of
$t^0$ defined by eq.~(\ref{tdef}) above satisfies properties t1 and t2.

To show that t1 holds, we note that the prescription for extending
$\tau^0_k$ clearly is local in the appropriate sense. However, it
is not immediately obvious that the prescription yields a covariant
extension $\tau_k$ in the sense required by t1 since the prescription
involves $\alpha_x$, whose definition requires, in addition to the
metric, a choice of a tetrad $e^a_\mu$ at $x$.  However, since any
other tetrad at $x$ is related by a Lorentz-transformation, it follows
immediately from the Lorentz invariance of the extensions ${u}$ in
\eqref{expan} that different choices of tetrad lead to the same
distribution $\tau_k$. It follows that each $\tau_k$ is locally
constructed from the metric in a covariant way in the sense required
by t1.

In order to see that $r_m$ is local and covariant in the sense of t1,
it is sufficient to show that $r_m[\psi^* \g]$ is equal to
$\psi^* r_m[\g]$ for any diffeomorphism $\psi$ on $U$. 
We already know that this is true off the total diagonal $\Delta_{n+1}$, 
as the unextended distribution $r^0$ has this
property. Thus, the difference between the two expressions must be a
distribution supported on the total diagonal. Moreover,
the scaling degree of this distribution must be less than $4n-1$, by 
our choice $m = d - 4n$. It is well known that there are no 
such distributions apart from the zero distribution (essentially 
because the delta function and its derivatives have scaling 
degree $\ge 4n$). Therefore the difference must in fact be zero, showing that
$r_m$ satisfies t1. Since all terms on the right side of
eq.~(\ref{tdef}) satisfy t1, it follows that $t$ satisfies this property.

To establish t2, we first show that the extensions $\tau_k[\g]$ have
an almost homogeneous scaling under rescalings of the metric in the
sense of t2. To see this, we consider a term $C \cdot \alpha_x^* u$ in
the expansion \eqref{expan}.  By Theorem 4.1, the curvature term $C$
will scale as $\lambda^{-k}$ under a rescaling of the metric by
$\lambda^2$.  On the other hand, for the term $\alpha_x^{*} u$, since
$\alpha_x$ is just the inverse of the exponential map at $x$, a
rescaling of the metric will correspond precisely to a coordinate
rescaling by a factor of $\lambda$ in the distributions $u$. By Lemma
\ref{help}, these distributions scale like $\lambda^{k-d}$ up to
logarithmic corrections under such a coordinate rescaling.  Therefore, each
individual term in formula \eqref{expan} for $\tau_k$ has an almost
homogeneous scaling with degree $d$ under rescalings of the metric. On
the other hand, the almost homogeneous scaling of $r_m$ under a
rescaling of the metric can be proven by a argument similar to the
proof that $r_m$ is local and covariant. Consequently, we see that $t$
satisfies property t2.

It also is relatively straightforward to prove that each $\tau_k$
occurring in eq.~\eqref{tdef} satisfies properties t3--t5. We know that
$\tau_k$ is a finite sum of terms of the form $C(x) \cdot \alpha_x^*
u(y)$, with $C(x)$ a polynomial in the curvature and its derivatives.
Since $C(x)$ is smooth in $x$, we have
\begin{multline}
\label{81}
\WF(C \cdot \alpha^* u) 
\subset 
\Big\{ (x, \textstyle{\sum} \left[
\tfrac{\partial \alpha_x}{\partial x} \right]^t k_i; 
\xi_1, \left[\tfrac{\partial \alpha_x}{\partial \xi_1} \right]^t
k_1; \dots;
\xi_n, \left[\tfrac{\partial \alpha_x}{\partial \xi_n} \right]^t
k_n ) \Big|\\
\left( \alpha_x(\xi_1), k_1; \dots;
\alpha_x(\xi_n), k_n\right) 
\in \WF(u) \Big\}.
\end{multline}
Here, we have written $y = (\xi_1, \dots, \xi_n)$ and each 
$\xi_i$ denotes a point in a convex normal neighborhood of 
$x$, and {\it not} the Riemannian normal coordinates 
of that point. (This makes a difference here, since we are 
considering variations in $x$.) In eq.~\eqref{81}, 
$\partial \alpha_x/\partial \xi_i$ denotes the matrix of partial
derivatives of $\alpha_x$ with respect to $\xi_i$ at fixed $x$, and
$\partial \alpha_x/\partial x$ denotes the matrix of partial
derivatives of $\alpha_x (\xi_i)$ with respect to $x$ at fixed
$\xi_i$. However, at $\xi_i = x$ we have $\partial 
\alpha_x(\xi_i)/\partial \xi_i =
-\partial \alpha_x(\xi_i)/\partial x$, since moving $\xi_i$ infinitesimally
away from $\xi_i = x$ has the same effect on $\alpha_x(\xi_i)$ as moving $x$
infinitesimally by the same amount in the opposite direction.  It
follows that if $(x, k_1; \dots; x, k_{n+1}) \in \WF(C \cdot \alpha^*
u)$, then $\sum k_i = 0$. This means precisely that $\WF(C \cdot
\alpha^* u) \restriction_{\Delta_{n+1}} \perp T(\Delta_{n+1})$, i.e.,
the microlocal spectral condition, t3, is satisfied. Similarly, by
using the fact that $C(x)$ is a polynomial in the curvature and
$\alpha_x$ is the inverse of the exponential map---so that both $C(x)$
and $\alpha_x$ have appropriate smooth and analytic dependence on the
metric---together with the fact that $u$ is independent of the metric,
we find that the smoothness (t4) and analyticity (t5) conditions are
satisfied by $\tau_k$.

Thus, we would be done if our expression (\ref{tdef}) for $t$
corresponded to a suitably convergent power series. However, as
already noted in remark (1) at the end of subsection 4.1, this is not the case,
i.e., the remainder term, $r_m$, in eq.~(\ref{tdef}) is not expected to
converge to zero in any sense useful for our purposes as $m
\rightarrow \infty$. Therefore, in order to prove that $t$ satisfies
properties t3--t5, it is necessary to explicitly analyze the
remainder term $r_m$. This is technically
quite cumbersome, since essentially the only thing useful that is
known about $r_m$ is that it is the extension to $\Delta_{n+1}$
defined by eq.~(\ref{seq}) of the expression $r^0_m$ given by
eqs.~(\ref{psidef}) and (\ref{psidef2}).  Equation~(\ref{seq})
expresses $r_m$ as a weak limit of distributions whose wave front set
properties are known, but wave front set properties are not preserved
under weak convergence, so we must show that the sequence (\ref{seq})
converges in a suitably strong sense to enable us to prove that $r_m$
satisfies properties t3--t5. This will be accomplished in the proof
of the following proposition, which---as will be explained in the
remark following the statement of the 
proposition---will complete the proof that the
extensions $t$ satisfy the properties t3--t5.

\begin{prop}
Let $\g^{(s)}$ be a smooth one-parameter family of smooth metrics, and
let $r_m(s,x_1,\dots,x_n)$ denote the remainder term in
eq.~\eqref{tdef}, viewed as a distribution on $\mr \times U^n$. (Here
$m = d - 4n$, where $d$ is the scaling dimension of $t$.) Then the
wave front set of $r_m$ satisfies
\begin{equation}
\label{rwfperp}
\WF(r_m) \restriction_{\mr \times \Delta_n} 
\perp T(\mr \times \Delta_n),   
\end{equation}
where the notation ``$\perp$'' was introduced below eq.~(\ref{perp}).
Similarly if $\g^{(s)}$ is an analytic one-parameter family of
analytic metrics, then \eqref{rwfperp} holds for the analytic wave
front set.
\end{prop}

\paragraph{Remark.} If we choose $\g^{(s)} = \g$ for all $s$, the above 
proposition implies that $r_m$ satisfies the microlocal spectral
condition t3. The proposition also implies that $r_m$ satisfies the
smoothness and analyticity conditions, t4 and t5. In fact, the
proposition asserts a somewhat stronger version of these
conditions, as it shows that the wave front set of
$r_m(s,x_1,\dots,x_n)$ not only cannot contain any points of the form
$(s,\rho;x,k_1;\dots;x,k_n)$ with $\sum k_i \neq 0$ but it also cannot
contain any such points with $\rho \neq 0$. Since each $\tau_k$ has
already been shown above to satisfy t3--t5, it follows that $t$
satisfies t3--t5 if $r_m$ does. Thus, our construction of time
ordered products satisfying properties T1--T9 of section 2 will be
completed once we have completed the proof of this proposition.

\begin{proof}
We will give the proof only for the analytic case; the proof for the
smooth case is similar, though somewhat simpler because the estimates
needed to establish the wave front set properties are simpler in nature
in the smooth case. As before, we proceed by induction in the number,
$n$, of variables $(x_1,\dots,x_n)$ on which $r_m$ depends. We
inductively assume that the analytic wave front set version of
eq.~(\ref{rwfperp}) holds for all $r_m$ that depend on $n$ or fewer
variables. By a slight generalization of the proof given above that
$\tau_k$ satisfies t3--t5, it can be shown that if $\g^{(s)}$ is an
analytic one-parameter family of analytic metrics then each $\tau_k$
also satisfies eq.(\ref{rwfperp}). Consequently our inductive hypothesis
implies that $t(s, x_1, \dots, x_n)
\equiv t[\g^{(s)}](x_1, \dots, x_n)$ satisfies
\begin{equation}
\label{wfperp}
\WF_A(t) \restriction_{\mr \times \Delta_n} 
\perp T(\mr \times \Delta_n),   
\end{equation}

From the distributional coefficients $t$ depending on 
$n$ or fewer spacetime arguments and the real parameter $s$ we obtain, 
by our inductive constructions, the distributional coefficients 
$t^0(s, x_1, \dots, x_{n+1}) \equiv t^0[\g^{(s)}](x_1, \dots, x_{n+1})$ 
depending on $n+1$ spacetime arguments and the real parameter $s$. 
These distributions are defined everywhere in $\mr \times U^{n+1}$, 
except for ($\mr$ times) the total diagonal, $\Delta_{n+1}$. 
Their analytic wave front set $\WF_A(t^0)$ is therefore a subset of 
$T^*(\mr \times (U^{n+1} \setminus \Delta_{n+1}))$.  
By essentially the same arguments as given in \cite[Sec. 7]{bf}
(modulo a straightforward modification of those arguments with 
regard to the  additional parameter $s$ on which the $t^0$ depend), 
the distribution $t^0(s, x_1, \dots, x_{n+1})$
can be expressed as a finite sum of terms of the form 
\begin{equation}
\label{t0decomp}
t(s, \{x_i\}_{i \in I}) t(s, \{x_j\}_{j \in I^c}) \prod_{i \in I, j \in I^c}
H_F(s, x_i, x_j)^{a_{ij}}.
\end{equation}
on each of the open sets $C_I$ introduced in eq.~\eqref{ci}.
Here, the $a_{ij}$ are certain natural numbers,  
$I$ is a nonempty proper subset of the set $\{1, \dots, n+1\}$ 
and  $I^c$ is its complement. (Note that since $I$ is a nonempty
proper subset, the expression~\eqref{t0decomp} only involves 
the distributional coefficients $t$ depending on $n$ or fewer
spacetime arguments.) Finally, 
$H_F(s, x_1, x_2) \equiv H_F[\g^{(s)}](x_1, x_2)$ is the local 
Feynman parametrix introduced below eq.~\eqref{HF}, for our 
analytic 1-parameter family of metrics. It can be seen by an 
explicit calculation that 
\begin{equation}
\WF_A(H_F) \restriction_{\mr \times \Delta_2} \perp T(\mr \times \Delta_2).
\label{WFHF}
\end{equation}
For each of the sets $I$ described above, let us define a projection map 
$\pi_I$ from $\mr \times U^{n + 1}$ to $\mr \times U^{|I|}$ 
(with $|I|$ the number of elements in $I$) by 
\begin{equation}
\pi_I:
(s, x_1, \dots, x_{n+1}) \to (s, \{x_i\}_{i \in I}).
\end{equation}
Using the rules for calculating 
the analytic wave front set of products of distributions~\cite{h}, 
we find from eq.~\eqref{t0decomp}
that the analytic wave front set of $t^0$ restricted to 
the open sets $\mr \times C_I$ is estimated by 
\begin{multline}
\WF_A(t^0) \restriction_{\mr \times C_I}
\subset 
(\pi_I^* \WF_A(t) \cup \{0\}) + (\pi_{I^c}^* \WF_A(t) \cup \{0\}) \\
+ \sum_{i\in I, j\in I^c} \sum^{a_{ij}} \,\,
(\pi_{\{i,j\}}^* \WF_A(H_F) \cup \{0\}) 
\subset T^*(\mr \times (U^{n+1} \setminus \Delta_{n+1})).
\end{multline}
If we now take the closure in $T^*(\mr \times U^{n+1})$ 
of the sets on both sides of the above relation (we denote this closure by an 
overbar), take the union over all $I$, 
and use eqs.~\eqref{wfperp} and~\eqref{WFHF}, then we obtain
\begin{equation}
\label{twfperp1}
\overline{\WF_A(t^0)} \restriction_{\mr \times \Delta_{n+1}} \perp
T(\mr \times \Delta_{n+1}).  
\end{equation}

From the properties of $\tau_k$, it then follows that
$r^0_m(s,x_1,\dots,x_{n+1})$ also satisfies the same condition, i.e.,
\begin{equation}
\label{wfperp1}
\overline{\WF_A(r^0_m)} \restriction_{\mr \times \Delta_{n+1}} \perp
T(\mr \times \Delta_{n+1}),
\end{equation}
Note that eq.~\eqref{wfperp1} imposes a nontrivial
restriction (beyond what we already know) on the wave front set of
$r^0_m$. Our aim is to show that \eqref{wfperp1} continues to hold
for the extension, $r_m$. In order to simplify the discussion, we will
show here only the weaker result that, for a fixed metric $\g$,
$r_m(x_1,\dots,x_{n+1})$ satisfies
\begin{equation}
\label{rwfperp2}
\WF_A(r_m) \restriction_{\Delta_{n+1}} 
\perp T(\Delta_{n+1}),   
\end{equation}
However, the arguments can be generalized straightforwardly to prove
\eqref{rwfperp} in $n+1$ variables for an analytic one-parameter
family of analytic metrics $\g^{(s)}$.

As above, we choose relative coordinates $(x, y)$ around the total
diagonal. We will identify the point $x \in M$ with its coordinates in
some chart, and we will identify $y=(\xi_1,\dots, \xi_n)$ with its
Riemannian normal coordinates relative to $x$,
so that the diagonal corresponds to $y = 0$. With this choice of
coordinates, we identify $t^0$ (and likewise, $r^0_m$) with a
distribution defined on $X \times (Y \setminus 0)$, where $X$ is an
open set in $\mr^4$ and $Y$ is an open neighborhood of the origin
$\mr^{4n}$. Let $x_0$ be some fixed point in $X$. It is possible to
construct a sequence of smooth functions of the form $\phi_N(x, y) =
\phi_N'(x) \phi_N''(y)$, where $\phi_N' \in C_0^\infty(K')$ is 1 in
a neighborhood of $x_0$, such that $\phi_N''$ vanishes in a
neighborhood of $0$ and is 1 outside some larger neighborhood
$K''$, and where $\phi_N$ satisfies the estimate 
\begin{equation}
\label{phicutoff}
|\partial^{\alpha+\beta} \phi_N|
\le C_\alpha^{|\beta|+1}(N+1)^{|\beta|} \quad
\forall |\beta| \le N = 1, 2, \dots. 
\end{equation}
If $f$ is a test function with support sufficiently close to $(x_0,
0)$, then the extension $r_m$ is defined by eq.~\eqref{seq}. For our
purposes, it is convenient to make the choice $\vartheta^{(j)} =
(\phi_{N})_{2^j}$, where the subscript $2^j$ means the pull-back by
the function $(x, y) \to (x, 2^j y)$.

In order to show $\WF_A(r_m) \perp T(\Delta_{n+1})$,
we must demonstrate that $(x_0, k_0, 0, p_0)$ is in 
the complement of $\WF_A(r_m)$ whenever $k_0 \neq 0$. It is 
not difficult to see that this will follow if we can show that 
it is possible to choose $K = K' \times K''$ so small that   
\begin{equation}
\label{82}
|\widehat{(\theta_{N})_{2^j}r^0_m}(k, p)| 
\le 2^{-j/2}C^{N+1}((N+1)/(|k|+|p|))^N 
\end{equation}
for all $(k,p)$ in some conic neighborhood $F$ of 
of $(k_0, p_0)$ and for all natural numbers $N$ and
$j$. Here, $\theta_N \in C^\infty_0(K)$ is 
the cutoff function defined by  $\phi_N(x, y) - \phi_N(x, 2y)$. 
Note that the support of $\theta_N$ does not intersect the 
submanifold $X \times \{0\}$, and that the sequence
of cutoff functions $\theta_N$ is again bounded in $\mathcal{E}'(K)$ 
and satisfies the estimate \eqref{phicutoff}.

In order to analyze the Fourier transform 
on the left side of \eqref{82}, we observe that 
\begin{equation}
\widehat{(\theta_{N})_{2^j}r^0_m}(k, p) = 2^{-4nj}
\widehat{\theta_N (r^0)_{2^{-j}}}(k, 2^{-j}p),  
\end{equation}
where $(r^0_m)_{2^{-j}}$ denotes the pull-back of the 
distribution $r^0_m$ by the map $(x,y) \to (x, 2^{-j}y)$, and 
where the factor $2^{-4nj}$ is due to the fact that 
$r^0_m$ transforms as a density. 
Recalling our choice $m=d-4n$, 
we can write the quantity on the right 
side of this equation as
\begin{equation}
\label{xxx}
2^{-j} \sum_l (j \ln 2)^l 
\widehat{\theta_N \psi^0_l}(2^{-j}, k, 2^{-j}p),  
\end{equation}
where the $\psi^0_l \in \cD'(\mr \times X \times (Y \setminus 0))$
were defined in eq.~\eqref{psidef}, and where the Fourier transform is
with respect to the variables $x$ and $y$. 

We now claim that for any closed conic set $\Gamma$ in $\mr \times
\mr^4 \times \mr^{4n}$ not containing elements of the form $(0, 0, p)$
there is a neighborhood $K_0 \subset \mr \times X \times Y$ of $(0,
x_0, 0)$ such that for all $l$
\begin{equation}
\WF_A(\psi^0_l) \cap (K_0 \times \Gamma) = \emptyset. 
\label{K0}
\end{equation}
To prove this, we decompose $\psi^0_l$ into simpler pieces, whose analytic wave
front set is either known by the induction process or can be
determined by elementary means. For this, we shall define a family of analytic 
metrics depending analytically 
on parameters $s \equiv (v, \mu, x) \in P_1 \times
P_2 \times P_3 \equiv P$, where $P_1$ is a small neighborhood of $1$
in $\mr$, $P_2$ is a small neighborhood of $0$ in $\mr$ and where
$P_3$ is a convex normal neighborhood in $M$ with respect to $\g$.
In order to define this family, let $\chi_{x, \mu}$ be the
diffeomorphism which shrinks the Riemannian normal coordinates $\xi$ of a
spacetime point about the point $x \in P_3$ by a factor of
$\mu$. In terms of this family of diffeomorphism, our family of
metrics is given by
\begin{equation}
\g^{(s)} = (v\mu)^{-2} \chi_{x, \mu}^* \g. 
\end{equation}
This is a real analytic family of analytic metrics (but with $s$ now
ranging over the 6-dimensional parameter space, $P$, rather than over
$\mr$). We have already established that the analytic 
wave front set of the distribution
$t^0$ on $P \times (U^{n+1} \setminus \Delta_{n+1})$ satisfies
\eqref{twfperp1} (with $\mr$ replaced by $P$).  In order to relate
$\psi^0_l$ to $t^0$, we let $R^{(m)}$ be the map from test functions on
$\mr$ to smooth functions on $\mr$ given by 
\begin{equation}
(R^{(m)} f)(\lambda) = \frac{1}{m!}
\int_0^1 (1-\mu)^{m} f^{(m+1)}(\lambda \mu) \, d\mu.  
\end{equation}    
Furthermore, we set
\begin{equation}
D^{(l)} = \frac{1}{l!}(v\partial/\partial v + d)^l. 
\end{equation}
Note that if $f$ is a smooth function on $\mr$ with compact support,
then we have ${\rm supp}(^t\! R^{(m)} f) \subset {\rm supp}(f)$, where
$^t\! R^{(m)}$ denotes the transpose of $R^{(m)}$. Thus $^t \! R^{(m)}$ has
proper support. It now follows straightforwardly from the definition
of $\psi^0_l$ that we can rewrite the action of the distributions
$\psi^0_l$ on test functions $f \in C^\infty_0(\mr \times (U^{n+1}
\setminus \Delta_{n+1}))$ as
\begin{equation}
\label{psiwrite}
\psi^0_l(f) = (\jmath^* t^0)\Big[(^t \!D^{(l)}_{v} \delta( \, \cdot\, - 1))
\otimes ((^t \!R^{(m)}_{\mu} \otimes 1_{x_1 \dots x_{n+1}}^{})f) \Big],  
\end{equation}
where the subscripts on the operators indicate on which 
of the variables $(v, \mu, x_1, \dots, x_{n+1})$ they act, and where 
$\jmath^* t^0$ denotes the pull back of $t^0$ by the 
analytic map 
\begin{equation}
\jmath: (v, \mu, x_1, \dots, x_{n+1} ) 
\to (v, \mu, x= x_1, x_1, \dots, x_{n+1}) \in P \times U^{n+1}.
\end{equation} 
The analytic wave front set of $\psi^0_l$ can now be estimated
from eq.~\eqref{psiwrite}
using our knowledge of the analytic wave front set of 
$t^0$, eq.~\eqref{twfperp1}, together with the 
rules for calculating the wave front set of a distribution 
under composition with distribution kernels \cite[Thm. 8.5.5]{h}, 
and under pull-back by analytic maps \cite[Thm. 8.5.1]{h}. For this, we only
need to know the following additional facts: (i) The action of an analytic 
partial differential operator, such as $D^{(l)}$, does not enlarge 
the analytic wave front set and (ii) the analytic wave front set 
of the distribution kernel of $R^{(m)}$ (viewed as a bidistribution
on $\mr \times \mr$) does not contain any elements of the 
form $(\lambda, 0; \mu, \rho)$. The first statement is proven in 
\cite[Thm. 8.4.7]{h}, and the second statement can be checked directly.

This information suffices to conclude from eq.~\eqref{psiwrite} that 
\begin{equation}
\label{000}
\overline{\WF_A(\psi^0_l)} \restriction_{\mr \times \Delta_{n+1}}
\perp T(\mr \times \Delta_{n+1}).  
\end{equation}
If $\mr \times (U^{n+1} \setminus
\Delta_{n+1})$ is identified with subset of 
$\mr \times X \times (Y \setminus 0)$ 
via the above choice of coordinates, then 
this means that 
\begin{equation}
\overline{\WF_A(\psi^0_l)} \restriction_{\mr \times X \times \{0\}} 
\perp T(\mr \times X \times  \{0\}).
\end{equation}
As a consequence, the open set $T^*(\mr \times X \times Y) 
\setminus \overline{\WF_A(\psi^0_l)}$ 
contains a set of the form $K_0 \times \Gamma$
as claimed in eq.~(\ref{K0}), provided that $K_0$ is chosen to be 
sufficiently sharply concentrated about the point $(0, x_0, 0)$.

We now blow up the sequence of cutoff functions $\theta_N \in
C^\infty_0(K)$ to a bounded sequence cutoff 
functions in $C^\infty_0(K_0)$ which still  
satisfy the inequality \eqref{phicutoff}, and which we shall denote by the 
same symbol. It then follows that with these cutoff functions, 
\begin{equation}
\label{---}
|\widetilde{\theta_N \psi^0_l}(\rho, k, p)| \le C^{N+1}((N+1)/(|\rho|+
|k|+|p|))^{N}
\end{equation} 
for all $(\rho, k, p) \in \Gamma$, provided the support $K_0$ of $\theta_N$
is sufficiently sharply concentrated near $(0, x_0, 0)$, where 
the tilde denotes the Fourier transform in $\mr \times \mr^4 \times \mr^{4n}$. 
With our new choice for $\theta_N$, we can write \eqref{xxx} as 
\begin{eqnarray}
\widehat{(\theta_{N})_{2^j}r^0_m}(k, p)
= (2\pi)^{-1/2} 2^{-j}
\sum_l (j\ln 2)^l \int_\mr \widetilde{\theta_N \psi^0_l}(\rho, k, 2^{-j} p)
e^{-\i 2^{-j} \rho} 
\,d\rho,  
\end{eqnarray}
where the sum is finite.
Now the cone $\Gamma$ can be chosen 
such that $(\rho, k, 2^{-j} p) \in \Gamma$ for all points $(k, p)$ in the 
cone $F$, all $\rho$ and all $j$. Thus we can use \eqref{---} to estimate 
\begin{equation}
|\widehat{(\theta_N)_{2^j}r^0_m}(k, p)|
\le 2^{-j/2} C^N(N/|k|)^{N-1}.
\end{equation}
For $(k, p)$ in the cone $F$ it holds 
that $|k| > \epsilon |p|$ for some $\epsilon > 0$. This enables
us to estimate the above expression further by 
\begin{equation}
\le 2^{-j/2} C^N(N/(|k| + |p|))^{N-1} 
\end{equation}
for all $(k, p)$ in $F$ and all natural numbers $N$ and $j$. This 
is what we wanted to show. 
\end{proof}

\paragraph{Remark.}

The distributions $t$ have now been shown to have an analytic
dependence on the metric in the sense of condition t5. This makes it 
possible to establish certain analyticity properties of the distributions
$u$ in the scaling expansion for the $t$, as we will now show. We 
know that if $(s, \rho; x_1, k_1; \dots; x_{n+1}, k_{n+1})$ is an element in
$\WF_A(t)$, then the element $(x_1, k_1; \dots; x_{n+1}, k_{n+1})$
must necessarily be in the set $\cC^{(s)}_T$, given by
eq.~\eqref{gamtdef}. By our scaling
expansion, we know that the distributions $u$ are given in terms of
$s$ derivatives of $t$ (evaluated at $s=0$), so we can calculate
$\WF_A(u)$ from $\WF_A(t)$ by the rules for the analytic wave front
set under restriction and differentiation. It is straightforward to
see that this gives
\begin{eqnarray}
\label{uawfs}
\WF_A(u) &\subset& \Big\{(\xi_1, k_1; \dots; \xi_n, k_n) 
\in T^* \mr^{4n} \setminus \{0\} \,\, \Big| \,\, 
\text{$\exists$ decorated graph $G(p)$ in $(\mr^4, \mbox{\boldmath $\eta$})$ with}
\nonumber \\
&& \text{vertices $0, \xi_1, \dots, \xi_n$ such that
$k_i = \sum_{e: s(e) = i} p_e - \sum_{e: t(e) = i} p_e \quad \forall i$}
\Big\}, 
\end{eqnarray}
where we use the graph-theoretical notation introduced in T3, and 
where $\mbox{\boldmath $\eta$}$ is the Minkowski metric.

\section{Conclusions and Outlook}

In this paper, we have given a construction of local, covariant time
ordered products of an arbitrary number of local Wick powers. These
local time ordered products were shown to satisfy properties T1--T9
of section 2. They therefore fulfill the assumptions of the uniqueness
theorem of our previous paper \cite[Thm. 5.2]{hw}. Consequently, for
any given polynomial order in the free field, any other prescription
for defining local time ordered products with the same properties will
differ from the prescription given in the present paper by products of
local curvature terms and lower order time ordered products, as
specified precisely in our uniqueness theorem. Although in this paper
we considered only a massless Klein-Gordon scalar field, our results
can be generalized straightforwardly to allow mass, and we do not
anticipate any difficulties in generalizing our results to 
fields with higher spin. Largely for notational simplicity, we also restricted
consideration to time ordered products of Wick powers that do not
contain derivatives of the field, but it should be straightforward to
generalize our construction to allow Wick powers of differentiated
fields (subject only to the caveat of footnote \ref{der}).

An important tool in our analysis was the scaling expansion introduced
in subsection 4.1 for the distributions $t$ appearing in the local
Wick expansion of time ordered products. In essence, this scaling
expansion gives corrections to the ``scaling limit'' of \cite{fh},
organized in powers of the curvature (and the dimensionful parameters,
if any are present). The scaling expansion generalizes to arbitrary
$t$ in the local Wick expansion
the usual ``short distance expansion'' for the Feynman propagator
(see remark (3) at the end of subsection 4.1). Although we restricted
consideration here to the distributions $t$, a similar scaling
expansion will exist for any local, covariant field which satisfies
appropriate wave front set and scaling properties. The properties of
the general scaling expansion for local covariant fields is currently
under investigation.

The results of this paper essentially complete the analysis of the
existence, uniqueness, and renormalizability of the perturbative
expansion of nonlinear quantum fields (with polynomial
self-interaction) in curved spacetime. It is natural to ask whether an
``exact'' formulation of nonlinear quantum field theory in curved
spacetime can be given. The Wightman axioms and other similar systems
cannot be straightforwardly generalized to curved spacetime on account
of their essential usage of Poincare invariance and the existence of a
preferred vacuum state. We are currently investigating the possibility
that the notion of a local, covariant quantum field (together with
suitable microlocal spectral conditions, etc.) may enable one to give
a useful formulation of axiomatic quantum field theory in curved
spacetime.

\medskip
\paragraph{Acknowledgements:} We would like to thank K.~Fredenhagen
and B.~S.~Kay for helpful discussions. In particular Kay's observation
that the leading order divergence of the ``fish graph'' (see remark
(3) at the end of subsection 4.1) could be renormalized by Minkowski
spacetime methods was useful to us with regard to our development of
the scaling expansion for general time ordered products. 
We thank Kay for making the results of his unpublished work~\cite{k}
available to us at an early stage of this work. This work
was supported in part by NSF grant PHY00-90138 to the University of
Chicago.

\appendix

\section{Smooth and analytic variation of distributions}

A key requirement that we impose on our definition of Wick polynomials
and their time ordered products is that they have appropriate
smooth/analytic dependence on the spacetime metric. The purpose of
this appendix is to elucidate the notion of smooth and analytic
variation of distributions.

To begin, let $X$ be a manifold and for each $s \in \mr$ let $u^{(s)}:
X \rightarrow \mc$ be a smooth (i.e., $C^\infty$) function. It is
useful to view $u^{(s)}$ as a map $u:\mr \times X \rightarrow \mc$. We
say that $u^{(s)}$ varies smoothly with $s$ if the map $u$ is
smooth. Note that this requirement of (joint) smoothness of the map
$u$ is stronger than the possible alternative requirement that
$u^{(s)}(x)$ be a smooth function of $s$ for each fixed $x \in
X$. This latter notion of (separate) smoothness in $s$ would not be a
natural one in the context of this paper for the following reason: We
consider one-parameter families of spacetimes $(M,\g^{(s)})$ and there
is no natural way of identifying spacetimes with different values of
$s$. However, the notion of separate smoothness is not invariant under
diffeomorphisms $\psi^{(s)}: X \rightarrow X$ that are (jointly)
smooth in $(s,x)$.

Now, for each $s \in \mr$ let $u^{(s)} \in \mathcal{D}'(X)$, i.e.,
$u^{(s)}$ is a distribution on $X$. We wish to define a notion of
smooth variation of $u^{(s)}$ with $s$ that corresponds to the notion
of (joint) smoothness of functions as defined in the previous
paragraph. To do so, it is useful to view $u^{(s)}$ as a distribution,
$u$, on $\mr \times X$. The basic idea of our definition is to require
$u$ to be ``not any more singular than each $u^{(s)}$''. One possible
way of implementing this notion would be to demand that the wave front
set of $u$ be contained in the wave front set of $u^{(s)}$ in the sense
that $\WF(u) \subset \{(s,\rho; x, k) \mid \rho=0, (x,k) \in
\WF(u^{(s)})\}$. However, this definition is unsatisfactory for the
following two independent reasons. First, the requirement that $\rho =
0$ is too strong in that it would, in particular, require the
singularities of $u^{(s)}$ to ``remain in a fixed location in $X$'' as
$s$ is varied. This would not be invariant under a one parameter
family of diffeomorphisms $\psi^{(s)} : X \rightarrow X$ that are
(jointly) smooth in $(s,x)$. It should be noted that the distributions,
$u^{(s)}$, of interest in this paper have singularities on the light
cones of $\g^{(s)}$ and, hence, their singularities cannot ``remain in
a fixed location'' for non-conformal variations of $\g$. Consequently,
we shall not require $\rho = 0$ in our definition. Second, if
$u^{(s)}$ happens to be ``less singular than normal'' for some value
of $s$, then under the above proposed definition, $u$ would fail to
vary smoothly with $s$ even if, in a naive sense, its variation with
$s$ was perfectly smooth. For example, $u^{(s)}(x) = s
\delta(x)$ would fail to be smooth at $s=0$ because $\WF(u^{(0)})
= \emptyset$ but $\WF(u)$ includes points with $s=0$. For this reason,
we will define a more general notion of smoothness with respect to an
arbitrary specified family of cones $\cC^{(s)}$. (Here, a {\it cone}
$\cC$ is a subset of $T^*X \backslash \{0\}$ having the property that
if $(x,k) \in \cC$, then $(x, \lambda k) \in \cC$ for all $\lambda >
0$.) For the definition to be nontrivial, we must choose $\cC^{(s)}$
so that $\WF(u^{(s)}) \subset \cC^{(s)}$, but we need not choose 
$\cC^{(s)} = \WF(u^{(s)})$.

\begin{defn}
\label{smoothdef} 
Let $u^{(s)}$ be a one-parameter family of distributions on a manifold
$X$ and let $\cC^{(s)}$ be a family of cones. We say that $u^{(s)}$
varies smoothly with $s$ with respect to $\cC^{(s)}$ if the wave front
set of the corresponding distribution $u$ on $\mr \times X$ satisfies
\begin{equation}
\WF(u) \subset \{(s,\rho; x,k) \in 
T^*(\mr \times X) \setminus \{0\} \mid (x,k) \in \cC^{(s)} \}
\label{smooth}
\end{equation}
\end{defn}

\paragraph{Remarks.}
(1) To illustrate the meaning of the above definition, let us consider
the two extreme case, namely (a) when the cones are trivial,
$\cC^{(s)} = \emptyset$, and (b) when the cones are maximal,
$\cC^{(s)} = T^*X \setminus \{0\}$. In the first case (a) we
immediately get that $\WF(u) = \emptyset$, so $u$ is smooth jointly in
$(s,x)$. In the second case (b), it might appear that our smoothness
condition is in fact empty. However, this is not the case, since
eq.~(\ref{smooth}) implies that no element of the form $(s, \rho; x,
0)$ can be in $\WF(u)$. Thus, for example, if $u = v \otimes \phi$
with $v$ a distribution on $\mr$ and $\phi$ a distribution on $X$, not
depending on $s$, eq.~(\ref{smooth}) requires that $v$ is smooth.

\smallskip
\noindent
(2) Let $u_1^{(s)}$ and $u_2^{(s)}$ be two families of distributions
which are smooth with respect to cones $\cC^{(s)}_1$ respectively
$\cC^{(s)}_2$. Then the rules for calculating the wave front set of a
sum of two distributions gives that the family $u_1^{(s)} + u_2^{(s)}$
is smooth with respect to the cones $\cC^{(s)}_1 \cup
\cC^{(s)}_2$. Likewise, if $\{0\} \notin \cC^{(s)}_1 + \cC^{(s)}_2$,
for each $s$, then the product $u_1^{(s)} u_2^{(s)}$ can be defined
for each $s$ and defines a distribution jointly in $(s, x)$. Moreover,
the rules for calculating the wave front set of the product of two
distributions gives that product family is smooth with respect to the
cones $\cC^{(s)}_1 + \cC^{(s)}_2$.
\medskip

The above definition allows us to define the notion of the smooth
variation of a one parameter family, $\omega^{(s)}$, of continuous {\it states}
on the algebras $\cW(M, \g^{(s)})$ of the spacetimes $(M,\g^{(s)})$:
We say that $\omega^{(s)}$ varies smoothly with $s$ if each of the
$n$-point functions ${\omega_n}^{(s)}$ of 
$\omega^{(s)}$---viewed as a distribution on
$M^n$---varies smoothly with $s$ in the sense of Definition
\ref{smoothdef} with $\cC^{(s)} = \WF(\omega^{(s)}_n)$. A one-parameter
family of {\it fields} $\Phi^{(s)}: \mathcal{D}(M^n) \rightarrow
\cW(M, \g^{(s)})$ in $n$ variables will then be said to vary smoothly
with $s$ with respect to the cones $\cC_\Phi^{(s)} \subset T^* M^n
\setminus \{0\}$, if the corresponding distributions $\omega^{(s)}
(\Phi^{(s)})$ vary smoothly with $s$ with respect to $\cC^{(s)}_\Phi$
for all smooth one-parameter families of states $\omega^{(s)}$. Since
continuous states on $\cW(M, \g)$ are precisely the Hadamard states whose truncated
$n$-point functions are smooth for $n \neq 2$ \cite{hr}, it follows
that $\Phi^{(s)}$ will be smooth if and only if $\omega^{(s)}
(\Phi^{(s)})$ is smooth for all smoothly varying families of 
Hadamard states $\omega^{(s)}$ with
smooth truncated $n$-point functions ($n\neq 2$).  This is the
requirement that we have adopted in condition T4 of section 2.

In \cite{hw}, a different notion of ``continuous variation'' of
$\Phi^{(s)}$ was introduced in the case where $\Phi^{(s)}$ is local
and covariant. Our uniqueness theorems for Wick polynomials and their
time ordered products used the hypothesis that they vary continuously
under smooth variations of the metric. It can be shown that our
requirement of smooth variation introduced here implies continuous
variation in the sense of \cite{hw}, so the uniqueness results of
\cite{hw} continue to hold under this replacement (as can also be
verified more straightforwardly by simply repeating the proofs with
the new hypothesis). It also can be shown that the construction of
local, covariant Wick polynomials given in \cite{hw} satisfies 
our new smoothness requirement with $\cC^{(s)}_\Phi = \emptyset$.

In order to define the notion of analytic variation of a one-parameter
family of distributions, $u^{(s)}$, on a real analytic manifold, $M$,
we first recall the definition of the analytic wave front set.
To begin, let $u$ be a function on $\mr^m$ which is a real analytic  
in a neighborhood of a point $x_0$ in $\mr^m$. 
Then it follows from Cauchy's integral formula, or rather its 
generalization to $\mc^m$, that 
\begin{equation}
|\partial^\alpha u| \le C^{|\alpha|+1}(|\alpha| + 1)^{|\alpha|} \quad
 \forall \alpha
\label{estim}
\end{equation}
in a neighborhood of $x_0$, where $C$ is some constant and multi-index
notation has been used. Conversely,
if the above estimate holds for a function $u$ in a neighborhood of
$x_0$, then $u$ is real analytic in that neighborhood. 

Condition \eqref{estim} can be formulated equivalently in terms of
Fourier transforms. Namely, one can show that an estimate of the form
\eqref{estim} holds if and only if there is a sequence $u_N$ of
compactly supported distributions equal to $u$ in some open ball
around $x_0$, which is bounded in the space $\mathcal{E}'(\mr^m)$ 
of distributions of compact support, and which satisfies
\begin{equation}
\label{estim1}
|\widehat u_N(k)| \le C^{N+1}((N+1)/|k|)^N \quad \forall N \in \mn.
\end{equation} 
This motivates the following definition. Let $u$ be a distribution on 
$X \subset \mr^m$. The analytic wave front set $\WF_A(u)$ is defined 
to be the complement of the set of all points $(x_0,k_0)$ in $X 
\times (\mr^m \setminus 0)$ such that there is an open neighborhood $U$
of $x_0$, a conic neighborhood $\Gamma$ of $k_0$ and a bounded 
sequence $u_N \in \mathcal{E}'(X)$ which is equal to $u$ on $U$ and
satisfies \eqref{estim1} whenever $k \in \Gamma$. 

It is clear from the definition that $u$ is given by a real analytic 
function in the neighborhood of points $x_0$ such that $\WF_A(u)$ 
contains no element of the form $(x_0, k_0)$.  
If $f: X \to Y$ is an analytic one-to-one map, then the analytic
wave front set of the pull-back $f^* u$ is given by 
$\{(x, df^t(x) k) \mid (f(x), k) \in \WF_A(u)\}$. This makes it possible, 
via localization in analytic charts,  
to define in an invariant way the analytic wave front set of a 
distribution on a real analytic manifold $X$.

In practice it is useful that $u_N$ can always be obtained as a product 
of $u$ and suitable cutoff functions, see \cite[Lem. 8.4.4]{h}: Let
$\WF_A(u) \cap (K \times F) = \emptyset$ for some compact subset 
$K$ of $X$ and some closed cone $F$, and let $\chi_N$ be a sequence
of cutoff functions in $C_0^\infty(K)$ such that for all $\alpha$  
\begin{equation}
\label{cutoff}
|\partial^{\alpha+\beta} \chi_N|
\le C_\alpha^{|\beta|+1}(N+1)^{|\beta|} \quad
\forall |\beta| \le N = 1, 2, \dots. 
\end{equation}
Then $u_N = \chi_N u$ is bounded in $\mathcal{E}'(X)$ and
satisfies \eqref{estim1} for all $k \in F$. 

A one-parameter family of distributions, $u^{(s)}$, on a real analytic
manifold, $X$, will be said to vary analytically with $s$ with respect
to the cones $\cC^{(s)}$ if eq.~\eqref{smooth} holds with $\WF$
replaced by $\WF_A$ everywhere in that equation. The notions of
analytic variations of states and fields can then be defined in
complete parallel with the definition of smooth variation given
above. This agrees with the notions previously introduced in
\cite{hw}.

\section{Properties of the distributions $u$ in the scaling expansion}

In the following proposition, we list 
some general properties which hold for any
almost homogeneous distribution on $\mr^m$.
The distributions $u$ in our scaling expansion are
particular examples of such distributions, and therefore
the proposition applies to them. In particular, combining the 
upper bound \eqref{uawfs} on the analytic wave front set 
with item (iii) in the proposition below, one can obtain
detailed information about the  analytic wave front set of the 
Fourier transforms $\widehat u$ of the distributions $u$ in the scaling 
expansion. This information suffices to establish that $\widehat u$ is
in fact an analytic function in a large portion of momentum-space, and 
that it is given by the boundary value of an analytic function for 
almost all momentum configurations (see \cite[Thm. 8.4.15]{h} 
for the appropriate criteria when a distribution can be 
written as the boundary value of an analytic function). 

\begin{prop}
Let $u \in \cD'(\mr^m)$ be an almost homogeneous distribution of 
degree $\rho$, i.e., $S^N_\rho u = 0$ for some $N \in \mn$, where
$S^N_\rho = (\sum y^i \partial/\partial y^i + \rho)^N$. Then 
\begin{enumerate}
\item[(i)]
$\WF_A(u) \subset \{(y, k) \in T^* \mr^m \setminus \{0\} 
\mid \sum y^i k_i = 0\}$. 
\item[(ii)]
$u$ can be extended to test functions in Schwartz space
and thereby defines a tempered distribution.  
\item[(iii)]
$\widehat u$ is again an almost 
homogeneous distribution, with degree $m-\rho$. 
Furthermore, we have   
\begin{eqnarray}
\label{uhawfs}
(x, k) \in \WF_A(u) &\Leftrightarrow& (k, -x) \in \WF_A(\widehat u) \quad
\text{if $x \neq 0, k \neq 0$,} \nonumber\\
x \in {\rm supp}(u) &\Leftrightarrow& (0, -x) \in \WF_A(\widehat u) \quad
\text{if $x \neq 0$,} \nonumber\\
k \in {\rm supp}(\widehat u) &\Leftrightarrow& (0, k) \in \WF_A(u)  \quad
\text{if $k \neq 0$.}
\end{eqnarray}
\end{enumerate}
\end{prop}
\begin{proof}
Since $S^N_\rho u = 0$, and since $S_\rho^N$ has analytic 
coefficients, we have by \cite[Thm. 8.6.1]{h} that 
\begin{equation}
\WF_A(u) \subset {\rm Char}(S_\rho^N)
= \{(y, k) \in T^* \mr^m \setminus \{0\} 
\mid \sum y^i k_i = 0\},  
\end{equation} 
where ${\rm Char}(P)$ is the characteristic set of a  
differential operator $P$, defined as the set of all 
$(y, k) \in T^*\mr^m \setminus \{0\}$ such that  
$p(y, k) = 0$, where $p$ is the principal symbol of $P$. 
This proves (i). 

Let $\chi_+$ 
and $\chi_-$ be smooth functions on $\mr$ with the property that 
$\chi_+ + \chi_- = 1$, and such that $\chi_-(r) = 0$ for
$r \le r_0$ and $\chi_-(r) = 1$ for $r \ge 2r_0$ for some $r_0 > 0$.  
We can therefore write
\begin{equation}
u(y) = \chi_+(|y|) u(y) + \chi_-(|y|) u(y). 
\end{equation}
The first distribution on the right side of this equation is
by definition of compact support, and therefore trivially a 
tempered distribution. Thus (ii) will follow if we can show 
that also the second distribution on the right side is tempered.
In order to prove this, we first show that it is 
possible to write $\chi_- u$ in ``polar coordinates''. 
For this, we note that $\WF(u) \cap N^*(\ms^{m-1}) 
= \emptyset$ by (i), which implies by \cite[Thm. 8.2.4]{h}
that $u$ has a well defined pull-back, $v$, to the unit sphere, 
$\ms^{m-1}$, in $\mr^m$. It follows from this, and the almost homogeneous
scaling of $u$ that it is possible to write
\begin{equation}
\label{polar}
u(\chi_- f) = \sum_{j=0}^{N-1} 
\int_0^\infty \int_{\ms^{m-1}} c_j\chi_-(r)r^{-\rho+m-1} \ln^j r \,
v(\hat y) f(r\hat y)
\,dr d\mu(\hat y), 
\end{equation}
where $(r,\hat y)$ denote polar coordinates in $\mr^m$, $d\mu$ is 
the standard measure on $\ms^{m-1}$ and the $c_j$ are unspecified 
complex constants. Since $v$ is a distribution
on $\ms^{m-1}$, there must exist differential operators 
$P_1, \dots, P_k$ on $\ms^{m-1}$ such that 
\begin{equation}
|v(h)| \le C \sum_{l \le k} \sup_{\hat y \in \ms^{m-1}} |P_l h(\hat y)| 
\end{equation}
for all test functions $h \in \cD(\ms^{m-1})$. Moreover
$\chi_-(r)r^{-\rho + m - 1}
\ln^j r$ is a smooth function on $\mr$ which grows polynomially together
with all its derivatives at infinity, and therefore is a 
tempered distribution. Combining these facts with eq.~\eqref{polar}, 
we easily get the estimate 
\begin{equation}
|u(\chi_- f)| \le C \sum_{|\alpha| \le a, 
|\beta| \le b} \sup_{y \in \mr^m} 
|y^\alpha \partial^\beta f(y)| 
\end{equation}
for some $a, b \in \mn$ and all $f$ in Schwartz space, 
thus showing that $\chi_- u$ is tempered. 

We come to the proof of (iii). That the Fourier transform of $u$
scales almost homogeneously with degree $m-\rho$ follows directly from
our definition. For the case of a distribution, $u$ that scales {\it
exactly} homogeneously with degree $\rho$, the remaining three
relations in eq.~\eqref{uhawfs} correspond precisely to
\cite[Thm. 8.4.18]{h}. The proof given in \cite[Thm. 8.4.18]{h} of the
second and third relations in~\eqref{uhawfs} can be applied without
modification to distributions with almost homogeneous scaling.  We
therefore only have to prove the first relation in~\eqref{uhawfs}.

Since the Fourier transform $\widehat u$ is again a tempered distribution
which scales homogeneously up to logarithmic terms, the problem is 
symmetric and it therefore suffices to show that 
\begin{equation}
(y_0, k_0) \notin \WF_A(u) \Rightarrow (k_0, -y_0) \notin \WF_A(\widehat u)
\end{equation}
if $y_0 \neq 0, k_0 \neq 0$. Choose 
compact neighborhoods $K$ and $\widehat K$ in 
$\mr^m \setminus 0$ of $y_0$ and $k_0$ such that
\begin{equation}
\WF_A(u) \cap (K \times \widehat K) = \emptyset, 
\end{equation}
and a sequence of cutoff functions 
$\chi_N \in C_0^\infty(\widehat K)$ such that 
\eqref{cutoff} is valid for every $\alpha$. We now estimate the 
Fourier transform of $v_N = \chi_N \widehat u$ in a conic 
neighborhood of $y_0$. By Fourier's inversion formula and the 
convolution theorem, we have 
\begin{equation}
\label{vndef}
\widehat v_N(-\lambda y) = \int_{\mr^m} u(x) \widehat \chi_N(x - \lambda y)
\, d^m x.  
\end{equation} 
We now estimate expression \eqref{vndef} for $|y - y_0| < r$ and arbitrary 
$\lambda$. For this, we consider two cases, first $0 < \lambda \le 1$, 
and second $\lambda > 1$. We begin with the first case. 
Since $u$ is a tempered distribution, we can estimate 
\begin{equation}
|\widehat v_N(-\lambda y)| \le C \sum_{|\alpha| \le a, |\beta| \le b}
\sup_{x \in \mr^m} |x^\alpha \partial^\beta \widehat \chi_N(x-\lambda y)|.
\end{equation}
Using \eqref{cutoff}, it is not difficult to estimate 
\begin{equation}
|y^\alpha \partial^\beta \widehat \chi_N(y)| \le 
C (N+1), \quad \forall N, |\alpha| \le a, |\beta| \le b
\end{equation}
From this one obtains the estimate 
\begin{equation}
|\widehat v_N(-\lambda y)| \le C (N+1) \le C_M^{N-M+1}((N - M + 1)/\lambda)
^{N-M}
\end{equation} 
for all $M < N$ and all $0 < \lambda \le 1$. 

In order to estimate \eqref{vndef} also for $\lambda > 1$, we 
use that $u$ scales almost homogeneous up to logarithms. 
This enables us to write
\begin{equation}
\label{penu}
\widehat v_N(-\lambda y) 
= \lambda^{\rho+m} \sum_{j\ge 0} \frac{\ln^j \lambda}{j!} 
\int_{\mr^m} u_j(x) \widehat \chi_N(\lambda(x - y))\, d^m x,   
\end{equation} 
where the sum is finite and where $u_j = S_\rho^j u$, with
$S_\rho = \sum y^i \partial/\partial y^i + \rho$. Since 
$S_\rho$ is a partial
differential operator with analytic coefficients, we conclude 
that $\WF_A(u_j)$ is not bigger than $\WF_A(u)$ 
and hence also has no intersection with 
$K \times \widehat K$. We are now in a position to use 
exactly the same arguments as in 
the proof of \cite[Thm. 8.4.18]{h} (modulo
a trivial additional estimate due to the logarithms), to show that there 
holds the estimate
\begin{equation}
|\widehat v_N(-\lambda y)| 
\le 
C^{N-M+1}((N - M + 1)/\lambda)^{N - M}, 
\quad \forall N, \lambda > 1, |y - y_0| < r, 
\end{equation}
for some natural number $M$. Together with \eqref{penu} this 
shows that we have  $\widehat v_N(y) 
\le C^{N-M+1}((N-M+1)/|y|)^{N-M}$ for all $y$ in the  
conic neighborhood 
\begin{equation}
\{ -\lambda y \in \mr^m \mid |y - y_0| \le r, \lambda > 0 \}
\end{equation} 
of $-y_0$ and for some fixed $M$. 
This proves the proposition.
\end{proof}

\end{document}